\documentclass[envcountsame,envcountsect]{llncs}
\usepackage{amsmath,amssymb,amstext,pdfsync,enumerate,gastex,braket}
\usepackage[disable,backgroundcolor=white,linecolor=black]{todonotes}
\usepackage{paralist}
\title{The monoid of queue actions}
\author{Martin Huschenbett\inst{1} \and Dietrich Kuske\inst{1} 
 \and Georg Zetzsche\inst{2}}
\institute{TU Ilmenau, Institut f\"ur Theoretische Informatik \and 
   TU Kaiserslautern, Fachbereich Informatik}

\newcommand{\cA}{{\mathcal A}}
\newcommand{\cB}{{\mathcal B}}

\newcommand{\cQ}{{\mathcal Q}}
\newcommand{\cR}{{\mathcal R}}
\newcommand{\cS}{{\mathcal S}}
\newcommand{\bN}{{\mathbb N}}
\renewcommand{\bar}{\overline}
\newcommand{\conjugate}{\overset{*}{\sim}}
\newcommand{\dual}{\delta}
\newcommand{\nfname}{\mathsf{nf}}
\newcommand{\nf}[1]{\nfname(#1)}
\newcommand{\NF}{\mathsf{NF}}

\newcommand{\describeImp}[2]{``(\ref{#1})$\Rightarrow$(\ref{#2})''}
\newcommand{\describeEq}[2]{``(\ref{#1})$\Leftrightarrow$(\ref{#2})''}

\newcommand{\NL}{\mathsf{NL}}
\newcommand{\nat}{\eta}
\newcommand{\ow}[1]{\mathsf{ow}(#1)}

\newcommand{\eshuffle}[2]{\left<#1,#2\right>}
\newcommand{\shuffle}[2]{\eshuffle{#1}{\bar{#2}}}

\newcommand{\OL}[2]{\mathsf{ol}(#1,#2)}

\newtheorem{observation}[theorem]{Observation}

\begin{document}
\maketitle

\begin{abstract}
We investigate the monoid of transformations that are induced by
sequences of writing to and reading from a queue storage.
We describe this monoid by means of a confluent and terminating
semi-Thue system and study some of its basic algebraic properties,
e.g., conjugacy. Moreover, we show that while several properties concerning
its rational subsets are undecidable, their uniform membership problem
is $\NL$-complete. Furthermore, we present an algebraic
characterization of this monoid's recognizable subsets. Finally, we prove
that it is not Thurston-automatic.
\end{abstract}

\section{Introduction}

Basic computing models differ in their storage mechanisms: there are finite
memory mechanisms, counters, blind counters, partially blind counters,
pushdowns, Turing tapes, queues and combinations of these mechanisms. Every
storage mechanism naturally comes with a set of basic actions like reading a
symbol from or writing a symbol to the pushdown. As a result, sequences of
basic actions transform the storage. The set of transformations induced by
sequences of basic actions then forms a monoid. As a consequence, fundamental
properties of a storage mechanism are mirrored by algebraic properties of the
induced monoid.  For example, the monoid induced by a deterministic finite
automaton is finite, a single blind counter induces the integers with addition,
and stacks induce polycyclic monoids~\cite{Kambites2009}.  In this paper, we
are interested in a queue as a storage mechanism. In particular, we investigate
the monoid $\cQ$ induced by a single queue.

The basic actions on a queue are writing the symbol $a$ into the queue
and reading the symbol $a$ from the queue (for each symbol $a$ from
the alphabet of the queue). Since $a$ can only be read from a queue if
it is the first entry in the queue, these actions are partial. Hence,
for every sequence of basic actions, there is a queue of shortest
length that can be transformed by the sequence without error (i.e.,
without attempting to read $a$ from a queue that does not start with
$a$). Our first main result (Theorem~\ref{T-semi-Thue}) in section \ref{S-semi-Thue} provides us
with a normal form for transformations induced by sequences of basic
actions: the transformation induced by a sequence of basic actions is
uniquely given by the subsequence of write actions, the subsequence
of read actions, and the length of the shortest queue that can be
transformed by the sequence without error. The proof is based on a
convergent finite semi-Thue system for the monoid~$\cQ$. In sections
\ref{S-definition} and~\ref{S-multiplication}, we derive equations
that hold in $\cQ$. The main result in this direction is
Theorem~\ref{T-multiplication}, which describes the normal form of the
product of two sequences of basic actions in normal form, i.e., it
describes the monoid operation in terms of normal forms.

Sections~\ref{S-conjugacy} and \ref{S-conjugators} concentrate on the
conjugacy problem in $\cQ$. The fundamental notion of conjugacy in
groups has been extended to monoids in two different ways: call $x$
and $y$ conjugate if the equation $xz=zy$ has a solution, and call
them transposed if there are $u$ and $v$ such that $x=uv$ and
$y=vu$. Then conjugacy $\approx$ is reflexive and transitive, but not
necessarily symmetric, and transposition $\sim$ is reflexive and
symmetric, but not necessarily transitive. These two relations have
been considered, e.g., in \cite{LenS69,Osi73,Ott84,Dub86,Zha91,Cho93}.
We prove that conjugacy is the transitive closure of transposition
and that two elements of $\cQ$ are conjugate if and only if their
subsequences of write and of read actions, respectively, are conjugate
in the free monoid. This characterization allows in particular to
decide conjugacy in polynomial time. In section~\ref{S-conjugators},
we prove that the set of solutions $z\in\cQ$ of $xz=zy$ is effectively
rational but not necessarily recognizable.

Section~\ref{S-rational} investigates algorithmic properties of rational
subsets of $\cQ$. Algorithmic aspects of rational subsets have received
increased attention in recent years; see~\cite{Lohrey2013survey} for a survey
on the membership problem.
Employing the fact that every element of $\cQ$ has only
polynomially many left factors, we can nondeterministically solve the
rational subset membership problem in logarithmic space. Since the
direct product of two free monoids embeds into $\cQ$, all the negative
results on rational transductions (cf.~\cite{Ber79}) as, e.g., the
undecidability of universality of a rational subset translate into our
setting (cf.\ Theorem~\ref{T-rational}). The subsequent
section~\ref{S-recognizable} characterizes the recognizable subsets of
$\cQ$. Recall that an element of $\cQ$ is completely given by its
subsequences of write and read actions, respectively, and the length
of the shortest queue that can be transformed without an
error. Regular conditions on the subsequences of write and read
actions, respectively, lead to recognizable sets in $\cQ$. Regarding
the shortest queue that can be transformed without error, the
situation is more complicated: the set of elements of $\cQ$ that
operate error-free on the empty queue is not recognizable. Using an
approximation of the length of the shortest queue, we obtain
recognizable subsets $\Omega_k\subseteq\cQ$. The announced
characterization then states that a subset of $\cQ$ is recognizable if
and only if it is a Boolean combination of regular conditions on the
subsequences of write and read actions, respectively, and sets
$\Omega_k$ (cf.~Theorem~\ref{thm:recognizability}).  In the final
section~\ref{S-automatic}, we prove that $\cQ$ is not automatic in the
sense of Thurston et al.~\cite{EPCHLT1992}\ (it cannot be automatic in
the sense of Khoussainov and Nerode \cite{KhoN95} since the free monoid with two generators is
interpretable in first order logic in $\cQ$).

\section{Preliminaries}

Let $A$ be an alphabet. As usual, the set of finite words over $A$, i.e. the free
monoid generated by $A$, is denoted $A^*$. Let $w = a_1 \dotso a_n \in A^*$ be some
word. The \emph{length} of $w$ is $|w| = n$. The word obtained from $w$ by reversing
the order of its symbols is $w^R = a_n \dotso a_1$. A word $u \in A^*$ is a
\emph{prefix} of $w$ if there is $v \in A^*$ such that $w = u v$. In this situation,
the word $v$ is unique and we refer to it by $u^{-1} w$. Similarly, $u$ is a \emph{suffix} of $w$
if $w = v u$ for some $v \in A^*$ and we then put $w u^{-1} = v$.
For $k \in \bN$, we let $A^{\leq k} = \set{ w \in A^* |{} |w| \leq k}$ and define $A^{>k}$ similarly.

Let $M$ be an arbitrary monoid. The \emph{concatenation} of two subsets $X,Y \subseteq M$
is defined as $X \cdot Y = \set{ xy | x \in X, y \in Y }$.
The \emph{Kleene iteration} of $X$ is the set
$X^* = \set{ x_1 \dotsm x_n | n \in \bN, x_1,\dotsc,x_n \in X }$.
In fact, $X^*$ is a submonoid of $M$, namely the smallest submonoid entirely including $X$.
Thus, $X^*$ is also called \emph{the submonoid generated by $X$}. The monoid $M$
is \emph{finitely generated}, if there is some finite subset $X \subseteq M$ such
that $M = X^*$.

A subset $L \subseteq M$ is called \emph{rational} if it can be
constructed from the finite subsets of $M$ using union, concatenation, and Kleene iteration only.
The subset $L$ is \emph{recognizable} if there are a finite
monoid $F$ and a morphism $\phi\colon M \to F$ such that $\phi^{-1}\left(\phi(L)\right) = L$.
The image of a rational set under a monoid morphism is again rational,
whereas recognizability is retained under preimages of morphisms.
It is well-known, that every recognizable subset of a finitely generated monoid
is rational. The converse implication is in general false.
However, if $M = A^*$ for some alphabet $A$, a subset $L \subseteq A^*$ is rational
if and only if it is recognizable. In this situation, we call $L$ \emph{regular}.

\section{Definition and basic equations}
\label{S-definition}

We want to model the behavior of a fifo-queue whose entries come from
a finite set~$A$ with $|A|\ge2$ (if $A$ is a singleton, the queue
degenerates into a partially blind counter). Consequently, the state of a
queue is an element from~$A^*$. The atomic actions are writing of the
symbol $a\in A$ into the queue (denoted $a$) and reading the symbol
$a\in A$ from the queue (denoted~$\bar a$). Formally, $\bar A$ is a
disjoint copy of $A$ whose elements are denoted $\bar a$. Furthermore,
we set $\Sigma=A\cup\bar A$. Then the atomic actions of the queue are
defined by the function $.\colon (A^*\cup\{\bot\})\times\Sigma^* \to
A^*\cup\{\bot\}$ as follows:
\begin{align*}
  q.\varepsilon & = q &
  q.au&= qa.u & q.\bar au &=
  \begin{cases}
    q'.u & \text{ if }q=aq'\\
    \bot & \text{ otherwise}
  \end{cases}
  & \bot.u &=\bot
\end{align*}
for $q\in A^*$, $a\in A$, and $u\in\Sigma^*$. Note that this means
that the free monoid $\Sigma^*$ acts on the set $A^*\cup\{\bot\}$.

\begin{example}
  Let the content of the queue be $q=ab$. Then $ab.\bar
  ac=b.c=bc.\varepsilon=bc$ and $ab.c\bar a=abc.\bar
  a=bc.\varepsilon=bc$, i.e., the sequences of basic actions $\bar ac$
  and $c\bar a$ behave the same on the queue $q=ab$. In
  Lemma~\ref{L-equations}, we will see that this is the case for any
  queue $q\in A^*\cup\{\bot\}$. Differently, we have $\varepsilon.\bar
  aa=\bot\neq \varepsilon=\varepsilon.a\bar a$, i.e., the sequences of
  basic actions $a\bar a$ and $\bar a a$ behave differently on certain
  queues.
\end{example}

\begin{definition}
  Two words $u,v\in\Sigma^*$ are \emph{equivalent} if $q.u=q.v$ for
  all queues $q\in A^*$. In that case, we write $u\equiv v$. The
  equivalence class wrt.\ $\equiv$ containing the word $u$ is denoted
  $[u]$.

  Since $\equiv$ is a congruence on the free monoid $\Sigma^*$, we can
  define the quotient monoid $\cQ=\Sigma^*/\mathord{\equiv}$
  and the natural epimorphism $\eta\colon \Sigma^* \to \cQ, u \mapsto [u]$.
  The monoid $\cQ$ is called the \emph{monoid of queue actions}.
\end{definition}

\noindent
Informally, the basic actions $a$ and $\bar{a}$ act ``dually'' on
$\Sigma^*\cup\{\bot\}$. We will see that this intuition can be made
formal based on the following definition: the map
$\dual\colon\Sigma^*\to\Sigma^*$ with $\dual(au)=\dual(u)\bar{a}$,
$\dual(\bar{a}u)=\dual(u)a$, and $\dual(\varepsilon)=\varepsilon$ for
$a\in A$ and $u\in \Sigma^*$ will be called the \emph{duality
  map}. Note that $\dual(uv)=\dual(v)\dual(u)$ and $\dual(\dual(u))=u$
(i.e., $\dual$ is an anti-morphism and an involution). We say the
equations $u\equiv v$ and $u'\equiv v'$ are \emph{dual} if
$u'=\delta(u)$ and $v'=\delta(v)$. In the following lemma, the
equations \eqref{eq:1} and \eqref{eq:2} are dual and the equation
\eqref{eq:3} is self-dual.

One consequence of Theorem~\ref{T-semi-Thue} below will be that dual
equations are equivalent. Nevertheless, before proving
Theorem~\ref{T-semi-Thue}, we have to prove dual equations separately.

\begin{lemma}\label{L-equations}
  Let $a,b\in A$. Then we have
  \begin{eqnarray}
    ab\bar b & \equiv & a\bar b b \label{eq:1}\\
    a\bar a \bar b & \equiv &\bar a a \bar b \label{eq:2}\\
    a\bar{b} & \equiv & \bar{b} a\text{ if }a\neq b\label{eq:3}\,.
  \end{eqnarray}
\end{lemma}

\noindent
From \eqref{eq:1} and \eqref{eq:3}, we get $ab\bar{c} \equiv
a\bar{c}b$ for any $a,b,c\in A$. Similarly, \eqref{eq:2} and
\eqref{eq:3} imply $a\bar{b}\bar{c} \equiv \bar{b} a \bar{c}$.

\begin{proof}
  Note that $a=b$ is not excluded. Suppose $qa=bq'\in bA^*$, then
  $q.ab\bar{b}=qab.\bar{b} =q'b$ and
  $q.a\bar{b}b=bq'.\bar{b}b=q'b$. Next let $qa\notin bA^*$ such that
  $qab\notin bA^*$. Then $q.ab\bar{b} =qab.\bar{b}=\bot$ and
  $q.a\bar{b}b=(qa.\bar{b}).b=\bot$. This finishes the proof of
  equation~\eqref{eq:1}.

  Let $q=aq'\in aA^*$.  Then $q.a\bar{a}\bar{b}=aq'a.\bar{a}\bar{b}
  =q'a.\bar{b}$ and $q.\bar{a}a\bar{b}=q'a.\bar{b}$. If
  $q=\varepsilon$ then
  $q.a\bar{a}\bar{b}=\bot=q.\bar{a}a\bar{b}$. Finally let
  $\varepsilon\neq q\notin aA^*$ such that $qa\notin aA^*$. Then
  $q.a\bar{a}\bar{b}=qa.\bar{a}\bar{b}=\bot$ and $q.\bar{a}a\bar{b}
  =\bot.a\bar{b}=\bot$. This finishes the proof of
  equation~\eqref{eq:2}.

  Suppose $a\neq b$. If $q=bq'\in bA^*$, then $q.a\bar{b} =
  qa.\bar{b}=q'a=q.\bar{b}a$. Next consider the case $q\notin
  bA^*$. Then $q.a\bar{b}=qa.\bar{b}=\bot$ since $qa\notin bA^*$ (the
  case $q=\varepsilon$ uses $a\neq b$). Similarly $q.\bar{b}a=\bot$
  since $q\notin bA^*$. Hence $a\bar{b} \equiv \bar{b}a$, i.e.,
  equation~\eqref{eq:3} holds.\qed  
\end{proof}

\noindent
Our computations in $\cQ$ will frequently make use of alternating
sequences of write- and read-operations on the queue. To simplify
notation, we define the shuffle of two words over $A$ and over
$\bar{A}$ as follows: Let $a_1,a_2,\dots,a_n,b_1,b_2,\dots,b_n\in A$
with $v=a_1a_2\dots a_n$ and $w=b_1b_2\dots b_n$. We write $\bar{w}$
for $\bar{b_1}\,\bar{b_2}\,\dots\,\bar{b_n}$ and  set 
\[
   \shuffle{v}{w} = a_1\bar{b_1}\,a_2\bar{b_2}\,\dots\,a_n\bar{b_n}
\]
(note that $\shuffle{v}{w}$ is only defined if $v$ and $w$ are
words over $A$ of equal length).

\begin{lemma}\label{L-rechnen}
  Let $u,v\in A^*$ and $a,b\in A$.
  \begin{enumerate}[(1)]
  \item If $|u|=|av|$, then $\shuffle{u}{av}\bar b\equiv \bar
    a\shuffle{u}{vb}$.
  \item If $|ub|=|v|$, then $a\shuffle{ub}{v} \equiv
    \shuffle{au}{v}b$.
  \item If $|u|=|v|$, then $a\shuffle{u}{v}\bar{b} \equiv\shuffle{au}{vb}$.
  \end{enumerate}
\end{lemma}

\noindent
We just remark that the equations in (1) and (2) are dual and that the
equation in (3) is self-dual.

\begin{proof}
  We prove the first claim by induction on the length of $v$ (that
  equals $|u|-1$): if $|v|=0$, then $u\in A$ and therefore
  $\shuffle{u}{av}\bar{b} = u\bar{a}\bar{b}\equiv
  \bar{a}u\bar{b}=\bar{a}\shuffle{u}{vb}$ by
  Lemma~\ref{L-equations}\eqref{eq:2}. Next let $|v|>0$. Then there exist
  $v_1,u_1\in A$ and $v_2,u_2\in A^*$ with $v=v_1v_2$ and
  $u=u_1u_2$. We obtain
  \begin{align*}
    \shuffle{u}{av} \bar{b}
    &= u_1\bar{a}\,\shuffle{u_2}{v_1v_2}\bar{b}\\
    &\equiv u_1\bar{a}\,\bar{v_1}\shuffle{u_2}{v_2b}
       &\qquad& \text{\ (by the induction hypothesis)}\\
    &\equiv \bar{a}u_1\,\bar{v_1}\shuffle{u_2}{v_2b}
       && \text{\ (by  Lemma~\ref{L-equations}\eqref{eq:2})}\\
    &=\bar{a}\shuffle{u}{vb}\,.
  \end{align*}
  This finishes the proof of the first claim, the second can be shown
  analogously.

  The third statement is trivial for $|v|=0$. If $|v|>0$, there are
  $v_1\in A$ and $v_2\in A^*$ with $v=v_1v_2$. Then we get from the
  first statement
  \begin{equation*}
  	\tag*{\qed}
  	a\shuffle{u}{v}\bar{b} \equiv a\bar{v_1}\shuffle{u}{v_2b} = \shuffle{au}{vb}\,.
  \end{equation*}
\end{proof}

\noindent
By induction on the length of $y$, one obtains the following
generalizations (for (2), induction on the length of $x$ is used).

\begin{proposition}\label{P-rechnen}
  Let $u,v,x,y,x',y'\in A^*$.
  \begin{enumerate}[(1)]
  \item if $xy=x'y'$ and $|x|=|y'|=|u|$, then $\shuffle{u}{x}\bar{y}
    \equiv \bar{x'}\shuffle{u}{y'}$.
  \item if $xy=x'y'$ and $|y|=|x'|=|v|$, then
    $x\shuffle{y}{v}\equiv\shuffle{x'}{v}y'$.
  \item If $|u|=|v|$ and $|x|=|y|$, then $x\shuffle{u}{v}\bar{y}
    \equiv\shuffle{xu}{vy}$. 
  \item If $|x| = |y|$, then
    $\shuffle{x}{y}\equiv x\bar{y}$.
  \end{enumerate}
\end{proposition}

\noindent
We note that, again, the equations in (1) and in (2) are dual and the
ones in (3) and (4) are self-dual. Moreover, (4) is a special case of (3)
for $u = v = \varepsilon$.

\begin{corollary}\label{cor:contextcommute}
  Let $u,v,w\in A^*$.
  \begin{enumerate}[(1)]
  \item If $|w| = |v|$, then $\bar{u}v\bar{w}\equiv v\bar{uw}$.
  \item If $|u| = |v|$, then $u\bar{v}w\equiv uw\bar{v}$.
  \end{enumerate}
\end{corollary}

\noindent
In this corollary, the second statement is the dual of the first.

\begin{proof}
  We prove the first claim. Let $u=b_1b_2\dots b_m$ and
  $w=b_{m+1}b_{m+2}\dots b_{m+n}$ with $b_i\in A$ for all $1\le i\le
  m+n$. Note that $n=|w|\ge|v|$. Then we have
  \begin{align*}
    \bar{u}v\bar{w}
      \tag{by Prop.~\ref{P-rechnen}~(3)}
      &\equiv \bar{b_1\dots b_m}\shuffle{v}{b_{m+1}\dots b_{m+|v|}} 
                 \bar{b_{m+|v|+1}\dots b_{m+n}} \\
      \tag{by Prop.~\ref{P-rechnen}~(1)}
      &\equiv \shuffle{v}{b_1\dots b_{|v|}}
              \bar{b_{|v|+1}\dots b_{m+n}} \\
      \tag{by Prop.~\ref{P-rechnen}~(3)}
      &\equiv v \bar{b_1\dots b_{|v|}}\,\bar{b_{|v|+1}\dots b_{m+n}} \\
      &=v\bar{u}\bar{w}\,.
  \end{align*}
  The second statement can be shown analogously.\qed
\end{proof}

\section{A semi-Thue system for $\cQ$}\label{S-semi-Thue}

We order the equations from Lemma~\ref{L-equations} as follows:
\begin{align*}
  a\bar{b} &\to \bar{b} a \text{ for }a\neq b\\
  ab\bar{b} & \to a\bar{b} b\\
  a\bar{a} \bar{b} & \to \bar{a} a \bar{b}
\end{align*}
Let $R$ be the semi-Thue system with the above three types of rules.
Note that a word over $\Sigma$ is irreducible if and only if it has
the form $\bar u\,\shuffle{v}{v}\,w$ for some $u,v,w\in A^*$. We find
it convenient to illustrate the irreducible word
$\bar{u}\shuffle{v}{v} w$ as follows:
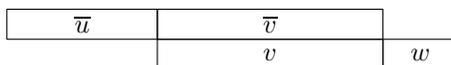
\begin{center}
  \begin{picture}(65,16) \gasset{Nmr=0}
    \node[Nw=20,Nh=4](baru)(10,10){$\bar u$}
    \node[Nw=30,Nh=4](barv)(35,10){$\bar v$}
    \node[Nw=30,Nh=4](v)(35,6){$v$} \node[Nw=10,Nh=4](w)(55,6){$w$}
  \end{picture}
\end{center}
Here, the blocks represent the words $\bar u$, $\bar v$, $v$, and $w$,
respectively where we placed the read-blocks (i.e., words over
$\bar{A}$) in the first line and write-blocks in the second. The
shuffle $\shuffle{v}{v}$ is illustrated by placing the corresponding
two blocks on top of each other.

\begin{lemma}\label{L-R-is-convergent}
  The semi-Thue system $R$ is terminating and confluent.
\end{lemma}

\begin{proof}
  We first show termination: For this, order the alphabet $\Sigma$ such that
  $\bar a < b$ for all $a,b\in A$. Then, for any rule $u\to v$ from $R$, the
  word~$v$ is length-lexicographically properly smaller than~$u$. Since the set
  $\Sigma^*$ ordered length-lexicographically is isomorphic to $(\bN,\le)$, the
  semi-Thue system~$R$ is terminating.

  To prove confluence of $R$, it suffices to show that $R$ is locally
  confluent. Note that the only overlap of two left-hand sides of $R$
  has the form $ab\bar{b}\bar{c}$ with $a,b,c\in A$. In this case, we can
  apply two rules (namely $ab\bar b\to a\bar bb$ and $b\bar b \bar
  c\to \bar bb\bar c$) which, in both cases, results in $a\bar bb\bar
  c$. \qed
\end{proof}

\noindent
Let $u\in\Sigma^*$. Since $R$ is terminating and confluent, there is a
unique irreducible word $\nf{u}$ with $u\xrightarrow{*}\nf{u}$. We
call $\nf{u}$ the \emph{normal form} of $u$ and denote the set of all
normal forms by $\NF \subseteq \Sigma^*$, i.e.,
\begin{equation*}
	\NF
	= \set{ \nf{u} | u \in \Sigma^* }
	= \bar{A}^* \set{ a \bar{a} | a \in A } A^* \,.
\end{equation*}
Note that, by
Lemma~\ref{L-equations}, we have $u\equiv\nf{u}$. Consequently,
$\nf{u}=\nf{v}$ implies $u\equiv v$ for any words $u,v\in\Sigma^*$. 
We next prove the converse implication.

\begin{lemma}\label{L-R-is-complete}
  Let $u,v\in\Sigma^*$ with $u\equiv v$. Then $\nf{u}=\nf{v}$.
\end{lemma}

\begin{proof}
  Let $\nf{u}=\bar{u_1}\shuffle{u_2}{u_2}u_3$ and
  $\nf{v}=\bar{v_1}\shuffle{v_2}{v_2}v_3$ and recall that
  $u\equiv\nf{u}\equiv\bar{u_1}u_2\bar{u_2}u_3$ holds by
  Prop.~\ref{P-rechnen}(3).  Hence, in the following, we can assume
  $u=\bar{u_1}u_2\bar{u_2}u_3$ and similarly
  $v=\bar{v_1}v_2\bar{v_2}v_3$.

  We first show $u_1=v_1$ by contradiction. So suppose $u_1\neq v_1$
  and, without loss of generality, $|u_1|\le |v_1|$. Then consider
  $q=u_1$. We get
  $q.u=\varepsilon.u_2\bar{u_2}\,u_3=u_3$. Furthermore,
  $u_1.\bar{v_1}=\bot$ since $u_1 \neq v_1$ and
  $|u_1|\le|v_1|$. Consequently
  $q.v=(q.v_1).v_2\bar{v_2}\,v_3=\bot$. Since this contradicts
  the assumption $q.u=q.v$, we obtain $u_1=v_1$.

  Without loss of generality, we can assume $|u_2|\le |v_2|$. Then we
  get
  \begin{align*}
    \bot&\neq u_2u_3 =u_1u_2.\bar{u_1} u_2\bar{u_2} u_3\\
     \tag{since $u \equiv v$}
     &= u_1u_2.\bar{v_1}{v_2}\bar{v_2} v_3 \\
     \tag{since $u_1 = v_2$}
     &= u_2.{v_2}\bar{v_2} v_3 \\
     &= (u_2.{v_2}\bar{v_2}).v_3\,.
  \end{align*}
  Hence $\bot\neq u_2.v_2\bar{v_2}=u_2v_2.\bar{v_2}$. It follows that $v_2$ is
  a prefix of $u_2v_2$ and, since $|u_2|\le |v_2|$, the word $u_2$ is a prefix
  of $v_2$. By contradiction, suppose $u_2$ is a proper prefix of $v_2$. Since
  $|A|\ge 2$, there exists $a\in A$ such that $u_2a$ is no prefix of $v_2$ (but
  still $|u_2a|\le|v_2|$). Then we get
  and
  \[
     u_1u_2a.u=u_1u_2a.\bar{u_1}{u_2}\bar{u_2}u_3=u_2a.{u_2}\bar{u_2}u_3=
     au_2u_3\neq\bot
  \]
  and
  \[
     u_1u_2a.v=u_1u_2a.\bar{v_1}{v_2}\bar{v_2}v_3=u_2a.{v_2}\bar{v_2}v_3=
     u_2av_2.\bar{v_2}v_3=\bot
  \]
  which contradicts the assumption $u\equiv v$. Hence $u_2=v_2$.

  To finally show $u_3=v_3$, consider the queue $q=u_1$. Then
  \begin{equation*}
  	\tag*{\qed}
    u_3=\varepsilon.{u_2}\bar{u_2}u_3=u_1.\bar{u_1}{u_2}\bar{u_2}u_3
      =u_1.\bar{v_1}{v_2}\bar{v_2}v_3
      =\varepsilon.{v_2}\bar{v_2}v_3
      = v_3\,.
  \end{equation*}
\end{proof}

\noindent
The above two lemmas ensure that $u\equiv v$ and $\nf{u}=\nf{v}$ are
equivalent. Hence, the mapping $\nfname\colon \Sigma^* \to \NF$
can be lifted to a mapping $\nfname\colon \cQ \to \NF$ by
defining $\nf{[u]} = \nf{u}$.

\begin{theorem}\label{T-semi-Thue}
  The natural epimorphism $\nat\colon\Sigma^*\to\cQ$ maps the
  set $\NF$ bijectively onto $\cQ$. The inverse of this bijection is the
  map $\nfname\colon \cQ \to \NF$.
\end{theorem}

\noindent
This theorem allows us to define projection maps on $\cQ$.  First, the
morphisms $\pi,\bar\pi\colon\Sigma^*\to A^*$ are defined by
$\pi(a)=\bar\pi(\bar{a})=a$ and $\pi(\bar{a})=\bar\pi(a)=\varepsilon$
for $a\in A$. In other words, $\pi$ is the projection of a word over
$\Sigma$ to its subword over $A$, and $\bar\pi$ is the projection to
its subword over $\bar{A}$, with all the
bars~$\bar{\rule[1ex]{1ex}{0pt}}$ deleted. E.g.,
$\pi(a\bar{b}\bar{a}b)=ab$ and $\bar\pi(a\bar{b}\bar{a}b)=ba$.  From
Theorem~\ref{T-semi-Thue}, we learn that $u\equiv v$ implies
$\pi(u)=\pi(v)$ and $\bar\pi(u)=\bar\pi(v)$. Hence, $\pi$ and
$\bar\pi$ can be lifted to morphisms $\pi,\bar\pi\colon\cQ\to A^*$
by $\pi([u])=\pi(u)$ and $\bar\pi([u])=\bar\pi(u)$.

Notice that the two projections $\pi(q)$ and $\bar{\pi}(q)$ of a queue action
$q \in \cQ$ do not entirely determine $q$, e.g., $[\bar{a} a] \not= [a \bar{a}]$.
However, in combination with the following property of $q$ they clearly do.

\begin{definition}
  Let $w \in \Sigma^*$ be a word and $\nf{w} = \bar{x} \shuffle{y}{y}
  z$ its normal form.  The \emph{overlap width} of $w$ and of $[w]$ is
  the number
\begin{equation*}
	\ow{w} = \ow{[w]} = |y| \,.
\end{equation*}
\end{definition}

\begin{observation}
\label{obs:complete_description}
Every $q \in \cQ$ is completely described by $\pi(q)$, $\bar{\pi}(q)$, and $\ow{q}$.
\end{observation}

\begin{remark}
  Let $q\in\cQ$ and $w = \nf{q}=\bar{x}\shuffle{y}{y}x$ its normal form.
  Then $x.\bar{x}\shuffle{y}{y}z =
  \varepsilon.\shuffle{y}{y}z = \varepsilon.z = z$, i.e., $q$
  transforms the queue $x$ without error. On the other hand, if $w$
  acts on a queue $x'$ without error, then $x$ is a prefix of
  $x'$. Hence $|x|$ is the length of the shortest queue that can be
  transformed by $q$ without error.
  Since $\ow{q} = |\pi(q)| - |x|$, $q$ is also uniquely
  given by $\pi(q)$, $\bar{\pi}(p)$, and the length
  of the shortest queue which is transformed by $q$ without error.
\end{remark}

\noindent
As announced before Lemma~\ref{L-equations}, we finally lift the
duality map $\delta$ from $\Sigma^*$ to~$\cQ$: Note that for any rule
$x\to y$ from the semi-Thue system, also $\dual(x)\to\dual(y)$ is a
rule. Therefore, if $u\overset{*}{\leftrightarrow} v$ for
$u,v\in\Sigma^*$, we also have
$\dual(u)\overset{*}{\leftrightarrow}\dual(v)$.  By
Theorem~\ref{T-semi-Thue}, this means $u\equiv v$ implies
$\dual(u)\equiv\dual(v)$.  Hence the lifted map $\dual\colon\cQ\to\cQ$
with $\dual([w])=[\dual(w)]$ is well-defined.  Observe that since
$\dual$ is an involution on $\Sigma^*$, it is also an involution on
$\cQ$ satisfying $\dual(xy)=\dual(y)\dual(x)$ for all $x,y\in\cQ$.

\section{Multiplication}
\label{S-multiplication}

For two words $u$ and $v$ in normal form, we want to determine the
normal form of $uv$. For this, the concept of \emph{overlap} of two
words will be important:

\begin{definition}
  For $u,v\in A^*$, let $\OL{v}{u}$ denote the longest suffix of $v$
  that is also a prefix of $u$.
\end{definition}

\begin{example}
$\OL{ab}{bc}=b$, $\OL{aba}{aba}=aba$, and $\OL{ab}{cba}=\varepsilon$.
\end{example}

\begin{lemma}\label{lemma:overlap}
  Let $u,v\in A^*$ with $|u|=|v|$ and set $s=\OL{v}{u}$, $r=vs^{-1}$
  and $t=s^{-1}u$. Then
  \[ 
    u\bar{v} \equiv \bar{r}\shuffle{s}{s}t\,.
  \]
\end{lemma}

\noindent
The equation $u\bar{v}\equiv \bar{r}\shuffle{s}{s}t$ can be visualized
as follows:
\begin{center}
  \begin{picture}(110,30) \gasset{Nmr=0}
    \node[Nw=30,Nh=5](baru)(45,25.5){$\bar v$}
    \node[Nw=30,Nh=5](v)(15,20.5){$u$}

    \node[Nframe=n](h)(10,8){$\equiv$}

    \node[Nw=10,Nh=5](baru)(25,10.5){$\bar {r}$}
    \node[Nw=20,Nh=5](barv)(40,10.5){$\bar {\OL{v}{u}}$}
    \node[Nw=20,Nh=5](v)(40,5.5){$\OL{v}{u}$}
    \node[Nw=10,Nh=5](w)(55,5.5){$t$}

    \node[Nframe=n](h)(65,8){$=$}

    \node[Nw=30,Nh=5](baru)(85,10.5){$\bar {v}$}
    \node[Nw=30,Nh=5](v)(95,5.5){$u$} 
  \end{picture}
\end{center}
In other words, when computing the normal form of $u\bar{v}$, all of
$\bar{v}$ except for the maximal suffix that is also a prefix of $u$
moves to the very beginning. The remaining suffix, i.e., $\OL{v}{u}$,
shuffles with the corresponding prefix, and the rest of $u$ moves to
the end.

\begin{proof}
  Let $u=a_1a_2\dots a_n$ and $v=b_1b_2\dots b_n$ with $a_i,b_i\in A$
  for all $1\le i\le n$. We prove the statement by induction on
  $n$. For $n=0$, the statement is trivial, so we may assume $n>0$.
  If $u=v$, we have $\OL{v}{u}=u$, confirming the equation. If $u\ne
  v$, there is some~$i\in\{1,2,\dots,n\}$ such that $a_i\ne b_i$.
  Then we have
  \begin{align*}
    \shuffle{u}{v}
      \tag{Lemma~\ref{L-equations}\eqref{eq:3}}
      &= \shuffle{a_1\dots a_{i-1}}{b_1\dots b_{i-1}}
         a_i\bar{b_i}
         \shuffle{a_{i+1}\dots a_n}{b_{i+1}\dots b_n}\\
      \tag{Lemma~\ref{L-rechnen}(1)}
      &\equiv \shuffle{a_1\dots a_{i-1}}{b_1\dots b_{i-1}}
         \bar{b_i} a_i
         \shuffle{a_{i+1}\dots a_n}{b_{i+1}\dots b_n} \\
      \tag{Lemma~\ref{L-rechnen}(2)}
      &\equiv \bar{b_1}\shuffle{a_1\dots a_{i-1}}{b_2\dots b_i}
         \shuffle{a_i\dots a_{n-1}}{b_{i+1}\dots b_n}a_n \\
      &= \bar{b_1}\shuffle{a_1\dots a_{n-1}}{b_2\dots b_n}a_n\,.
  \end{align*}
  Let $u'=a_1\cdots a_{n-1}$ and $v'=b_2\cdots b_n$. Then the
  induction hypothesis guarantees
  \[ 
     \shuffle{u'}{v'}\equiv
    \bar{r'}\shuffle{s'}{s'}t'\ 
      \text{for}\ s'=\OL{v'}{u'},\ r'=v's'^{-1},\ t'=s'^{-1}u'\,. 
  \]
  Consequently, we have
  \[
    \shuffle{u}{v}
    \equiv
    \bar{b_1}\,\bar{r'}\shuffle{s'}{s'}t'\,a_n\,.
  \]
  Since $u\ne v$ and $|u|=|v|$, we have $\OL{v}{u}=\OL{v'}{u'}$ and
  hence $s=s'$.  This means $b_1r'=b_1v's^{-1}=vs^{-1}=r$ and
  $t'a_n=s^{-1}u'a_n=s^{-1}u=t$.  Thus,
  \begin{equation*}
  	\tag*{\qed}
     \shuffle{u}{v}
     \equiv \bar{b_1}\bar{r'}\shuffle{s}{s}t' a_n
     = \bar r\shuffle{s}{s}t\,.
  \end{equation*}•
\end{proof}

\noindent
We next show that the above lemma holds even without the assumption
$|u|=|v|$.

\begin{lemma}\label{lemma:puremult}
  Let $u,v\in A^*$ and set $s=\OL{v}{u}$, $r=vs^{-1}$ and
  $t=s^{-1}u$. Then
  \[ 
    u\bar{v} \equiv \bar{r}\shuffle{s}{s}t\,.
  \]
\end{lemma}

\begin{proof}
  First, we assume $|u|\le |v|$ and write $v=xy$ with $|y|=|u|$. Then
  Corollary~\ref{cor:contextcommute} yields $u\bar{v}=u\bar{xy}\equiv
  \bar{x}u\bar{y}$ and by Lemma \ref{lemma:overlap}, we have
  %Let $v=b_1\cdots b_m$ with $b_i\in A$
  %for all $1\le i\le m$. Applying Prop.~\ref{P-rechnen}(1) twice, we
  %obtain
  %\begin{align*}
  %  u\bar{v} 
  %  &\equiv \bar{b_1\dots b_{m-|u|}}\shuffle{u}{b_{m-|u|+1}\dots b_m}\\
  %  &\equiv \bar{b_1\dots b_{m-|u|}}\, u\, \bar{b_{m-|u|+1}\dots b_m}\,.
  %\end{align*}
  %Set $x=b_1\dots b_{m-|u|}$ and $y=b_{m-|u|+1}\dots b_m$ such that
  %$u\bar{v}\equiv \bar{x}u\bar{y}$.
  \[ 
    u\bar{y} \equiv \bar{r'}\shuffle{s'}{s'}t'\ 
    \text{for}\ s'=\OL{y}{u},\ r'=ys'^{-1},\ t=s'^{-1}u\,.
  \]
  Since $|u|=|y|$, we have $s=\OL{xy}{u}=\OL{y}{u}=s'$. Furthermore,
  $xr'=xys'^{-1}=vs^{-1}=r$ and $t'=s'^{-1}u=s^{-1}u=t$. Hence
  \[
     u\bar{v}\equiv \bar{x}u\bar{y} 
             \equiv \bar{x}\,\bar{r'}\shuffle{s'}{s'}t'
             = \bar{r}\shuffle{s}{s}t
  \]
  is the desired equality.

  The case $|u|>|v|$ is handled by duality: define $s=\OL{u^R}{v^R}$,
  $r=u^R s^{-1}$, and $t=s^{-1}v^R$. Then, by what we showed above,
  $u\bar{v} =\dual(v^R\bar{u^R}) = \dual(\bar{r}\shuffle{s}{s}t) =
  \bar{t^R}\shuffle{s^R}{s^R}r^R$. Note that $s^R=\OL{v}{u}$,
  $r^R={s^R}^{-1}u$, and $t^R=v {s^R}^{-1}$.\qed
\end{proof}

\noindent
Finally, we describe the normal form of the product of two words in
normal form. In other words, we describe the multiplication of $\cQ$
in terms of words in normal form.

\begin{theorem}\label{T-multiplication}
  Let $u_1,u_2,u_3,v_1,v_2,v_3\in A^*$ and set
  $s=\OL{u_2v_1v_2}{u_2u_3v_2}$, $r=u_2v_1v_2s^{-1}$, and
  $t=s^{-1}u_2u_3v_2$. Then
  \[ 
    \bar{u_1}\shuffle{u_2}{u_2}u_3\cdot
     \bar{v_1}\shuffle{v_2}{v_2}v_3
    \equiv
     \bar{u_1r}\shuffle{s}{s}tv_3\,.
   \] 
\end{theorem}

\noindent
This theorem can also be visualized:
\begin{center}
  \begin{picture}(110,50)(0,-20)
    \gasset{Nmr=0}
    \node[Nw=10,Nh=5](baru)(5,25.5){$\bar {u_1}$}
    \node[Nw=20,Nh=5](barv)(20,25.5){$\bar {u_2}$}
    \node[Nw=20,Nh=5](v)(20,20.5){$u_2$} 
    \node[Nw=10,Nh=5](w)(35,20.5){$u_3$}

    \node[Nframe=n](h)(45,22.5){$\cdot$}

    \node[Nw=10,Nh=5](baru)(55,25.5){$\bar {v_1}$}
    \node[Nw=20,Nh=5](barv)(70,25.5){$\bar {v_2}$}
    \node[Nw=20,Nh=5](v)(70,20.5){$v_2$} 
    \node[Nw=10,Nh=5](w)(85,20.5){$v_3$}

    \node[Nframe=n](h)(10,7.5){$\equiv$}

    \node[Nw=10,Nh=5](baru)(20,10.5){$\bar {u_1}$}
    \node[Nw=15,Nh=5](r)(32.5,10.5){$\bar {r}$}
    \node[Nw=35,Nh=5](barv)(57.5,10.5){$\bar {\OL{u_2v_1v_2}{u_2u_3v_2}}$}
    \node[Nw=35,Nh=5](v)(57.5,5.5){${\OL{u_2v_1v_2}{u_2u_3v_2}}$}
    \node[Nw=15,Nh=5](w)(82.5,5.5){$t$}
    \node[Nw=10,Nh=5](w)(95,5.5){$v_3$}

    \node[Nframe=n](h)(10,-12.5){$=$}

    \node[Nw=10,Nh=5](baru)(20,-9.5){$\bar {u_1}$}
    \node[Nw=50,Nh=5](barv)(50,-9.5){$\bar{u_2v_1v_2}$}
    \node[Nw=50,Nh=5](v)(65,-14.5){$u_2u_3v_2$}
    \node[Nw=10,Nh=5](w)(95,-14.5){$v_3$}
  \end{picture}
\end{center}

\begin{proof}
  We have
  \begin{align*}
    \bar{u_1}\shuffle{u_2}{u_2}u_3\cdot \bar{v_1}\shuffle{v_2}{v_2}v_3
     \tag{Prop.~\ref{P-rechnen}(3)}
     &\equiv
    \bar{u_1}u_2\bar{u_2}u_3 \bar{v_1}v_2\bar{v_2}v_3 \\
     \tag{Cor.~\ref{cor:contextcommute}}
     &\equiv
    \bar{u_1}u_2u_3\bar{u_2} \bar{v_1}v_2\bar{v_2}v_3 \\
     \tag{Cor.~\ref{cor:contextcommute}}
     &\equiv
    \bar{u_1}u_2u_3v_2\bar{u_2} \bar{v_1}\bar{v_2}v_3 \\
     \tag*{(Lemma \ref{lemma:puremult}) \qed}
     &\equiv
     \bar{u_1}\,\bar{r}\shuffle{s}{s} t v_3 \,.
  \end{align*}
\end{proof}

\noindent
As a consequence of Theorems~\ref{T-semi-Thue} and
\ref{T-multiplication}, we can show that the queue-monoid with two
letters contains all other queue-monoids as submonoids.

\begin{corollary}
  Let $\cQ_n$ be the queue-monoid defined by an alphabet with $n$
  letters. Then $\cQ_n$ embeds into $\cQ_2$ for any $n\in\bN$.
\end{corollary}

\begin{proof}
  Let $\cQ_n$ be generated by the set $A=\{\alpha_i\mid 1\le i\le n\}$
  and let $\cQ_2$ be generated by $B=\{a,b\}$. Then define a
  morphism $\phi\colon (A\cup\bar A)^*\to (B\cup\bar B)^*$ by
  $\phi(\alpha_i)=a^{n+i}ba^{n-i}b$ and
  $\phi(\bar{\alpha_i})=\bar{a^{n+i}ba^{n-i}b}$. If $1\le i,j\le n$
  are distinct, then no non-empty suffix of $\phi(\alpha_i)$ is a
  suffix of $\phi(\alpha_j)$, i.e.,
  $\OL{\phi(\alpha_i)}{\phi(\alpha_j})=\varepsilon$. Hence
  $\phi(\alpha_i\bar{\alpha_j})\equiv \phi(\bar{\alpha_j}\alpha_i)$ by
  Theorem~\ref{T-multiplication}. Furthermore note that all the words
  $\phi(\alpha_i)$ have length $2n+2$. Consequently, by
  Cor.~\ref{cor:contextcommute}, we have
  $\phi(\alpha_i\alpha_j\bar{\alpha_j})\equiv
  \phi(\alpha_i\bar{\alpha_j}\alpha_j)$ and
  $\phi(\alpha_i\bar{\alpha_i}\bar{\alpha_j})\equiv
  \phi(\bar{\alpha_i}\alpha_i\bar{\alpha_j})$ for all $1\le i,j\le
  n$. From these observations and Lemma~\ref{L-equations}, we get
  $\phi(U)\equiv \phi(U')$ for all $U,U'\in (A\cup\bar A)^*$ with
  $U\equiv U'$.

  We next want to prove the converse implication. So let $U,U'\in
  (A\cup\bar A)^*$ with $\phi(U)\equiv\phi(U')$. There exist
  $\beta_i,\beta_i',\gamma_i,\gamma_i',\delta_i,\delta_i'\in A$ such
  that
  \begin{align*}
    \nf{U} &=
        \bar{\beta_1\dots\beta_k}\,
        \gamma_1\bar{\gamma_1}\dots\gamma_\ell\bar{\gamma_\ell}\,
        \delta_1\dots\delta_m\\\text{ and }
     \nf{U'} &=
        \bar{\beta'_1\dots\beta'_{k'}}\,
        \gamma'_1\bar{\gamma'_1}\dots\gamma'_{\ell'}\bar{\gamma'_{\ell'}}\,
        \delta'_1\dots\delta'_{m'}\,.
  \end{align*}
  Since $|\phi(\gamma_i)|=|\phi(\bar\gamma_i)|$, we obtain
  \begin{align*}
    \phi(U)&\equiv\phi(\nf{U})= \phi(\bar{\beta_1\dots\beta_k})\,
    \phi(\gamma_1\bar{\gamma_1})\dots\phi(\gamma_\ell\bar{\gamma_\ell})\,
    \phi(\delta_1\dots\delta_m) \\
      &\equiv
    \phi(\bar{\beta_1\dots\beta_k})\,
    \shuffle{\phi(\gamma_1)}{\phi(\gamma_1)}
          \dots\shuffle{\phi(\gamma_\ell)}{\phi(\gamma_\ell)}\,
    \phi(\delta_1\dots\delta_m)\\
      &=
    \phi(\bar{\beta_1\dots\beta_k})\,
    \shuffle{\phi(\gamma_1\dots\gamma_\ell)}{\phi(\gamma_1\dots\gamma_\ell)}\,
    \phi(\delta_1\dots\delta_m)\\
    \intertext{and similarly}
    \phi(U')&\equiv
    \phi(\bar{\beta'_1\dots\beta'_{k'}})\,
    \shuffle{\phi(\gamma'_1\dots\gamma'_{\ell'})}
            {\phi(\gamma'_1\dots\gamma'_{\ell'})}\,
    \phi(\delta'_1\dots\delta'_{m'})\,.
  \end{align*}
  Since $\phi(U)\equiv\phi(U')$, Theorem~\ref{T-semi-Thue} implies
  \begin{align*}
    \phi(\beta_1\dots\beta_k) &= \phi(\beta'_1\dots\beta'_{k'})\,,\\
    \phi(\gamma_1\dots\gamma_\ell) &= \phi(\gamma'_1\dots\gamma'_{\ell'})\,,\\
    \text{and }
    \phi(\delta_1\dots\delta_m) &= \phi(\delta'_1\dots\delta'_{m'})\,.
  \end{align*}
  Since $\phi$ acts injectively on $A^*$, this implies
  \begin{align*}
    \beta_1\dots\beta_k &= \beta'_1\dots\beta'_{k'}\,,\\
    \gamma_1\dots\gamma_\ell &= \gamma'_1\dots\gamma'_{\ell'}\,,\\
    \text{and }
    \delta_1\dots\delta_m &= \delta'_1\dots\delta'_{m'}
  \end{align*}
  and therefore $U\equiv\nf{U}=\nf{U'}\equiv U'$.

  In other words, the morphism $\phi\colon A^*\to B^*$ can be
  lifted to an injective morphism from $\cQ_n$ to $\cQ_2$, i.e.,
  $\cQ_n$ embeds into $\cQ_2$.\qed
\end{proof}

\section{Conjugacy}
\label{S-conjugacy}

In this section, we consider the relations of conjugacy and transposition
in the monoid of queue actions $\cQ$.

\begin{definition}
Let $M$ be a monoid and $p,q \in M$. Then $p$ and $q$ are \emph{conjugate},
in symbols $p \approx q$, if there exists $x \in M$ such that $p x = x q$.
Furthermore, $p$ and $q$ are \emph{transposed}, in symbols $p \sim q$,
if there are $x,y \in M$ with $p = x y$ and $q = y x$.
\end{definition}

\noindent
Observe that $\approx$ is reflexive and transitive whereas $\sim$ is reflexive
and symmetric. If $M$ is actually a group, then both relations coincide and
are equivalence relations, called conjugacy.
The same is true for free
monoids \cite[Prop.~1.3.4]{Lothaire1983} and special monoids \cite{Zha91},
but there are monoids where none of
this holds. In this section, we prove for the monoid~$\cQ$ that $\approx$ is
the transitive and reflexive closure of $\sim$, which is denoted by $\conjugate$.
Moreover, we give a simple (polynomial-time) characterization of
when $p \approx q$ holds.

Notice that the relation $\sim$ on $\cQ$ is self-dual in the following sense:
Let $p,q \in \cQ$ with $p \sim q$ and $x,y \in \cQ$ such that $p = x y$ and
$q = y x$. Then $\dual(p) = \dual(y) \dual(x)$ and $\dual(q) = \dual(x) \dual(y)$,
i.e., $\dual(p) \sim \dual(q)$.
Conversely, $\dual(p) \sim \dual(q)$ also implies $p \sim q$ because
$\dual$ is an involution. Consequently, $\conjugate$ is self-dual in the same sense as well.

\begin{lemma}\label{L-rrightunequal}
  Let $x,y\in A^*$ and $a\in A$. If $x\ne ya$, then $[\bar{x}ya]
  \conjugate [\bar{x}ay]$.
\end{lemma}

\begin{proof}
  If $x=\varepsilon$, we have
  $[\bar{x}ya]=[y][a]\sim[a][y]=[\bar{x}ay]$. Hence, let $x=ub$ with
  $u\in A^*$ and $b\in A$. If $b \not= a$, then 
  \[
    [\bar{u}\bar{b}\,ya]\sim[a\bar{u}\bar{b} y] = [\bar{u}a\bar{b}y]
    =[\bar{u}\bar{b}ay] \,.
  \]
  Henceforth, assume $b = a$. Thus,
  $x=ua$ and consequently $x\ne ya$ implies $u\neq y$.
  With $w = \nf{y \bar{u}}$, we have
  \[
    [\bar{x}ya]=[\bar{ua} ya] \sim [ya\bar{ua}] = [y\bar{u}a\bar{a}]
    = [w a \bar{a}] \,.
  \]
  Notice that $w$ cannot start with a write symbol and
  end with a read symbol at the same time, because this would imply
  $w \in \set{ a \bar{a} | a \in A}^*$ and hence $u = y$.
  On the one hand, if $w$ starts with a read symbol, we have
  \begin{equation*}
  	[w a \bar{a}] \sim [a \bar{a} w] = [\bar{a} a w] = [\bar{a} a y \bar{u}]
	\sim [\bar{u a} a y] = [\bar{x} y a] \,.
  \end{equation*}
  On the other hand, if $w$ ends with a write symbol, we obtain
  \begin{equation*}
	\tag*{\qed}
	[w a \bar{a}] = [w \bar{a} a] = [y \bar{u a} a] \sim [\bar{ua}ay]
	= [\bar{x}ay] \,.
  \end{equation*}
\end{proof}

\begin{lemma}\label{rotate}
For $x,y\in A^*$ and $a\in A$, we have
\begin{inparaenum}[(1)]
\item $[\bar{x}ya]\conjugate [\bar{x}ay]$ and
\item $[\bar{x a} y] \conjugate [\bar{a x} y]$.
\end{inparaenum}
\end{lemma}

\begin{proof}
  We show claim~(1) first.
  The case $x\ne ya$ was treated in Lemma \ref{L-rrightunequal} and we may
  therefore assume $x=ya$. Let $u=a_1a_2\dots a_k$ with
  $a_1,\dots,a_k\in A$ be the shortest nonempty prefix of $x$ such
  that $x=vu=uv$ for the complementary suffix $v\in A^*$.

  Then $x\ne a_{\ell+1}a_{\ell+2}\dots a_k\, v\, a_1a_2\dots a_\ell$
  for all $1\le \ell<k$ and hence, applying
  Lemma~\ref{L-rrightunequal} $k-1$ times, we get
  \begin{align*}
    [\bar{x} ay] & =[\bar{x} a_k va_1a_2\dots a_{k-1}]\\
    & \conjugate [\bar{x} a_{k-1}a_k va_1a_2\dots a_{k-2}]\\
    & \conjugate [\bar{x} a_{k-2} a_{k-1}a_k va_1a_2\dots a_{k-3}]\\
    & \hspace{0.5em} \vdots\\    
    & \conjugate [\bar{x}a_1\dots a_kv]    = [\bar{x} ya] \,.
  \end{align*}

Concerning the claim~(2), we first observe that
\begin{equation*}
	\dual([\bar{xa}y])
	= \left[\bar{y^R}ax^R\right]
	\conjugate \left[\bar{y^R}x^Ra\right]
	= \dual([\bar{ax}y]) \,.
\end{equation*}
Since $\sim$ is self-dual, we may conclude $[\bar{x a} y] \conjugate [\bar{a x} y]$.
\qed
\end{proof}

The announced description of $\approx$ is a characterization in terms of the
projections of the elements.
\begin{theorem}
  For any $p,q\in\cQ$, the following are equivalent:
  \begin{enumerate}[(1)]
  \item\label{cond:switch} $p\conjugate q$.
  \item\label{cond:exists} $p\approx q$.
  \item\label{cond:existsrev} $q\approx p$.
  \item\label{cond:projections} $\pi(p)\sim \pi(q)$ and
    $\bar{\pi}(p)\sim\bar{\pi}(q)$.
  \end{enumerate}
\end{theorem}

\begin{proof}
  If $p\sim q$ with $p=rs$ and $q=sr$, then $pr=rsr=rq$ and hence
  $p\approx q$. Since $\approx$ is transitive, this ensures
  \describeImp{cond:switch}{cond:exists}.

  In order to show
  \describeImp{cond:exists}{cond:projections},
  suppose $px=xq$. Then we have $\pi(p)\pi(x)=\pi(x)\pi(q)$ and
  $\bar\pi(p)\bar\pi(x)=\bar\pi(x)\bar\pi(q)$.  Since $\sim$ and
  $\approx$ coincide on the free monoid, this implies
  $\pi(p)\sim\pi(q)$ and $\bar\pi(p)\sim\bar\pi(q)$ and therefore
  \eqref{cond:projections}.

  Next, we prove
  \describeImp{cond:projections}{cond:switch}. So
  assume $\pi(p) \sim \pi(q)$ and $\bar\pi(p) \sim \bar\pi(q)$. There
  are unique words $r,s,t,u,v,w\in A^*$ with
  $p=[\bar{r}\shuffle{s}{s}t]$ and $q=[\bar{u}\shuffle{v}{v}w]$.  Note
  that $ts\sim st=\pi(p)\sim\pi(q)=vw\sim wv$ and
  $rs=\bar\pi(p)\sim\bar\pi(q)=uv$. Then we get
  \begin{align*}
    p &= [\bar{r}\shuffle{s}{s} t]\\
      \tag{$[s]\cdot[\bar{r}\bar{s}]=[\bar{r}\shuffle{s}{s}]$ by Theorem~\ref{T-multiplication}}
      &= [s\bar{r}\bar{s} t] \\
      &\sim [\bar{r}\bar{s} ts]\\
      \tag{$ts \sim wv$ and repeated application of Lemma~\ref{rotate}(1)}
      &\conjugate [\bar{r}\bar{s} wv] \\
      \tag{$rs\sim uv$ and repeated application of Lemma~\ref{rotate}(2)}
      &\conjugate [\bar{u}\bar{v} wv] \\
      &\sim [v\bar{u}\bar{v} w] \\
      \tag{$[v]\cdot[\bar{u}\bar{v}]=[\bar{u}\shuffle{v}{v}]$ by Theorem~\ref{T-multiplication}}
      &= [\bar{u}\shuffle{v}{v}w] \\
      &=q\,.
  \end{align*}
  Thus, we proved the equivalence of \eqref{cond:switch},
  \eqref{cond:exists}, and \eqref{cond:projections}. It follows in
  particular that $\approx$ is symmetric. Hence, \eqref{cond:exists}
  and \eqref{cond:existsrev} are equivalent as well.\qed
\end{proof}

\noindent
Given two words $u$ and $v$ over $\Sigma$, one can decide in quadratic
time whether $\pi(u)\sim \pi(v)$ and
$\bar\pi(u)\sim\bar\pi(v)$. Consequently, it is decidable in
polynomial time whether $[u]\approx[v]$ holds.

\section{Conjugators}
\label{S-conjugators}

\newcommand{\Conj}[1]{C(#1)}
\newcommand{\OverConj}[1]{D(#1)}

\begin{definition}
Let $M$ be a monoid and $x,y \in M$. An element $z \in M$ is a
\emph{conjugator of $x$ and $y$} if $xz = zy$. The set of all
conjugators of $x$ and $y$ is denoted
\begin{equation*}
	\Conj{x,y} = \set{ z \in M | xz = zy } \,.
\end{equation*}
\end{definition}

\noindent
Suppose that $M$ is a free monoid $A^*$ and consider $x,y \in A^*$.
It is well-known that $z \in A^*$ is a conjugator of $x$ and $y$ precisely if
there are $u,v \in A^*$ such that $x = uv$, $y = vu$, and $z \in u(vu)^*$.
Consequently, $\Conj{x,y}$ is a finite union of sets of the form $u(vu)^*$ and
hence regular. In contrast, Observation~\ref{obs:conjugators_not_recognizable}
and Theorem~\ref{thm:conjugators_rational} demonstrate
that in the monoid $\cQ$ sets of conjugators are always rational but in general
not recognizable.

\begin{observation}
\label{obs:conjugators_not_recognizable}
Let $a \in A$. The set $\Conj{[\bar{a}],[\bar{a}]}$ is not recognizable.
\end{observation}

\begin{proof}
We show the claim by establishing the equation
\begin{equation*}
	\nat^{-1}\left(\Conj{[\bar{a}],[\bar{a}]}\right) \cap a^* \bar{a}^*
	= \Set{ a^k \bar{a}^\ell | k,\ell \in \bN, k \leq \ell } \,.
\end{equation*}
To this end, consider $k,\ell \in \bN$ and let $z = \left[a^k \bar{a}^\ell\right]$.
On the one hand, if $k \leq \ell$, then
\begin{equation*}
	\nf{\left[\bar{a}\right] z}
	= \bar{a}^{\ell+1-k} (a \bar{a})^k
	= \nf{z \left[\bar{a}\right]} \,,
\end{equation*}
i.e., $z \in \Conj{[\bar{a}],[\bar{a}]}$.
On the other hand, if $k > \ell$, then
\begin{equation*}
	\nf{\left[\bar{a}\right] z}
	= \bar{a} (a \bar{a})^\ell a^{k-\ell}
	\not= (a \bar{a})^{\ell+1} a^{k-\ell-1}
	= \nf{z \left[\bar{a}\right]} \,,
\end{equation*}
i.e., $z \not\in \Conj{[\bar{a}],[\bar{a}]}$.
\qed
\end{proof}

\begin{theorem}
\label{thm:conjugators_rational}
Let $x,y \in \cQ$. Then the set $\Conj{x,y}$ is rational.
\end{theorem}

\noindent
The proof needs some preparatory lemmas and follows at the end of this section.
Throughout, we fix two elements $x,y \in \cQ$ as well as their
normal forms $\nf{x} = \bar{x_1} \shuffle{x_2}{x_2} x_3$ and
$\nf{y} = \bar{y_1} \shuffle{y_2}{y_2} y_3$.
Applying the projections $\pi$ and $\bar{\pi}$ to the equation $xz = zy$
for any $z \in \Conj{x,y}$ yields that $\pi(z)$ is a conjugator of $\pi(x)$ and $\pi(y)$
as well as that $\bar{\pi}(z)$ is a conjugator of $\bar{\pi}(x)$ and $\bar{\pi}(y)$.
Thus, the set
\begin{equation*}
	\OverConj{x,y}
	= \set{ z \in \cQ | \pi(xz) = \pi(zy) \mathrel{\&} \bar{\pi}(xz) = \bar{\pi}(zy) }
	\supseteq \Conj{x,y}
\end{equation*}
can be regarded as an overestimation of $\Conj{x,y}$. Recall that any $q \in \cQ$ is
completely determined by $\pi(q)$, $\bar{\pi}(q)$, and $\ow{q}$.
Thus, $z \in \OverConj{x,y}$ satisfies $z \in \Conj{x,y}$ if and only if
$\ow{xz} = \ow{zy}$. The proof of Theorem~\ref{thm:conjugators_rational}
basically exploits this observation in combination with the fact that the set
$D(x,y)$ can be rephrased as
\begin{equation*}
	\OverConj{x,y}
	= \pi^{-1}\bigl(\Conj{\pi(x),\pi(y)}\bigr) \cap
		\bar{\pi}^{-1}\bigl(\Conj{\bar{\pi}(x),\bar{\pi}(y)}\bigr)
\end{equation*}
and is hence recognizable.

\begin{lemma}
\label{lemma:overconj_ow_bounded}
Every $z \in \OverConj{x,y}$ satisfies $0 \leq \ow{xz} - \ow{z} \leq |\pi(x)|$.
\end{lemma}

\begin{proof}
Let $\nf{z} = \bar{z_1} \shuffle{z_2}{z_2} z_3$.
By Theorem~\ref{T-multiplication}, we have
\begin{equation*}
	\ow{xz} = |\OL{x_2 z_1 z_2}{x_2 x_3 z_2}| \leq |x_2 x_3 z_2| = |\pi(x)| + \ow{z} \,.
\end{equation*}
This proves the second inequation.

Since $\pi(z) \in \Conj{\pi(x),\pi(y)}$, we can apply the characterization of
conjugators in free monoids and write $\pi(x)=uv$ and $z_2z_3=\pi(z)=(uv)^ku$
for some $u,v\in A^*$  and $k\in\bN$.  Hence, $z_2=(uv)^\ell w$ for some prefix
$w$ of $uv$ and $\ell\in\bN$.  Thus, $z_2$ is a prefix of $x_2 x_3 z_2 = (u
v)^{\ell+1} w$ as well as a suffix of $x_2 z_1 z_2$.  Again by
Theorem~\ref{T-multiplication}, this implies $\ow{xz} \geq |z_2| = \ow{z}$,
i.e., the first inequation.  \qed
\end{proof}

\begin{lemma}
\label{lemma:overconj_ow_recognizable}
For every $k \in \bN$, the following set is regular:
\begin{equation*}
	G_k = \set{ \nf{z} | z \in \OverConj{x,y}, \ow{xz} - \ow{z} \geq k } \,.
\end{equation*}
\end{lemma}

\begin{proof}
Consider some $z \in \OverConj{x,y}$ and let $\nf{z} = \bar{z_1} \shuffle{z_2}{z_2} z_3$
be its normal form. Due to Theorem~\ref{T-multiplication}, we have $\ow{xz} \geq \ow{z} + k$
if and only if there is some $w \in A^*$ with $|w| \geq \ow{z} + k$ that is
a suffix of $x_2 z_1 z_2$ as well as a prefix of $x_2 x_3 z_2$.
Since $\ow{z} = |z_2|$, this is true precisely if there is some suffix $u \in A^{\geq k}$
of $x_2 z_1$ such that $u z_2$ is a prefix of $x_2 x_3 z_2$. According to
Lemma~\ref{lemma:overconj_ow_bounded}, any such $u$ also satisfies $|u| \leq |\pi(x)|$.
Altogether, this amounts to
\begin{equation*}
	G_k
	= \nat^{-1}\left(\OverConj{x,y}\right) \cap
		\bigcup_{\substack{u \in A^* \\ k \leq |u| \leq |\pi(x)|}}
			\bar{X_u} \, \phi(Y_u) \, A^* \,,
\end{equation*}
where $\phi\colon A^* \to \Sigma^*$ is the morphism defined by $\phi(v) = \shuffle{v}{v}$,
\begin{align*}
	X_u &= \set{ z_1 \in A^* | \text{$u$ is a suffix of $x_2 z_1$} } \,, \\
\intertext{and}
	Y_u &= \set{ z_2 \in A^* | \text{$u z_2$ is a prefix of $x_2 x_3 z_2$} } \,.
\end{align*}
Since $\OverConj{x,y}$ is recognizable, it suffices to show that $X_u$ and $Y_u$ are
regular for each $u \in A^*$ in order to prove the claim of the lemma.

Concerning $X_u$, observe that $u$ is a suffix of $x_2 z_1$ if $u$ is a suffix of $z_1$
or there is a factorization $u = v z_1$ of $u$ such that $v$ is a suffix of $x_2$. Thus,
\begin{equation*}
	X_u = A^* u \cup \set{ z_1 | v,z_1 \in A^*, u = v z_1, \text{$v$ is a suffix of $x_2$} }
\end{equation*}
and this set is clearly regular. Concerning $Y_u$, we first observe that
$Y_u = \emptyset$ if $u$ is not a prefix of $x_2 x_3$ and $Y_u = A^*$ if $u = x_2 x_3$.
If $u$ is a proper prefix of $x_2 x_3$, say $x_2 x_3 = u v$, then $Y_u$ is the set of all
$z_2 \in A^*$ such that $z_2$ is a prefix of $v z_2$. It is well-known that this is
precisely the prefix closure of $v^*$.
In each of these three cases, $Y_u$ is regular.
\qed
\end{proof}

\begin{proof}[of Theorem~\ref{thm:conjugators_rational}]
Consider some $k \in \bN$.
By Lemma~\ref{lemma:overconj_ow_recognizable}, the set
\begin{equation*}
	E_k
	= \set{ \nf{z} | z \in \OverConj{x,y}, \ow{xz} - \ow{z} = k }
	= G_k \setminus G_{k+1}
\end{equation*}
is regular. Our first goal is to show that the set
\begin{equation*}
	F_k = \set{ \nf{z} | z \in \OverConj{x,y}, \ow{zy} - \ow{z} = k }
\end{equation*}
is regular as well. To that end, it suffices to show that
$\dual(F_k)$ is regular because $\dual$ is an involution that preserves
regularity of subsets of $\Sigma^*$.

It is a matter of routine to check that $z \in \OverConj{x,y}$ holds true
precisely if ${\dual(z) \in \OverConj{\dual(y),\dual(x)}}$.
Since $\dual$ preserves the overlap width and $\dual(\nf{z}) = \nf{\dual(z)}$,
we thus obtain
\begin{equation*}
	\dual(F_k)
	= \set{ \nf{\dual(z)} |
		\dual(z) \in \OverConj{\dual(y),\dual(x)}, \ow{\dual(y)\dual(z)} - \ow{\dual(z)} = k } \,.
\end{equation*}
Using once more that $\dual$ is an involution, and hence surjective, yields
\begin{equation*}
	\dual(F_k)
	= \set{ \nf{z} |
		z \in \OverConj{\dual(y),\dual(x)}, \ow{\dual(y)z} - \ow{z} = k } \,. \\
\end{equation*}
Since the regularity of $E_k$ does not depend on the specific choice of $x$ and $y$,
this set and hence also $F_k$ are regular.

Recall that $z \in \OverConj{x,y}$ satisfies $z \in \Conj{x,y}$ precisely if
$\ow{xz} = \ow{zy}$. Using Lemma~\ref{lemma:overconj_ow_bounded}, we thus obtain
\begin{equation*}
	\set{ \nf{z} | z \in \Conj{x,y} }
	= \bigcup_{0 \leq k \leq |\pi(x)|} E_k \cap F_k \,.
\end{equation*}
Since this set is regular, $\Conj{x,y}$ is rational.
\qed
\end{proof}

\section{Rational subsets}
\label{S-rational}

This section studies decision problems concerning rational subsets of
$\cQ$.  While most of these problems are undecidable, the uniform
membership in rational subsets is $\NL$-complete.

Let $w\in\Sigma^*$. Then one can show that the number of left-divisors
of $[w]$ in $\cQ$ is at most $|w|^3$. This allows to define a DFA with
$|w|^3$ many states that accepts $[w]=\{u\in\Sigma^*\mid u\equiv
w\}$. The following lemma strengthens this observation by showing that
such a DFA can be constructed in logarithmic space.

\begin{lemma}\label{L-Automat-fuer-AeK}
  From $w\in\Sigma^*$, one can construct in logarithmic space a DFA
  accepting $[w]$.
\end{lemma}

\begin{proof}
  Let $w=a_1a_2\dots a_n$. For $1\le i\le n$ and $0\le j\le n$, we
  define $w[i,j]=a_ia_{i+1}\dots a_j$, in particular
  $w[i,j]=\varepsilon$ if $i>j$.

  Let $i,j,k,\ell\in\{0,1,\dots,n\}$ be natural numbers. For the
  quadrupel $p=(i,j,k,\ell)$, we define four words
  $p_1,p_2,p_2',p_3\in A^*$ setting
  \begin{itemize}
  \item $p_1=\bar\pi(w[1,i])$ and $p_2=\bar\pi(w[i+1,j])$ as well
    as
  \item $p_2'=\pi(w[1,k])$ and $p_3=\pi(w[k+1,\ell])$.
  \end{itemize}
  Then $p$ is a state of the DFA if and only if
  \begin{itemize}
  \item $p_2=p_2'$,
  \item $i=0$ or $a_i\in \bar{A}$ and similarly $j=0$ or
    $a_j\in\bar{A}$, and
  \item $k=0$ or $a_k\in A$ and similarly $\ell=0$ or $a_\ell\in A$.
  \end{itemize}
  Hence every state $p$ of the DFA stands for a word $u_p=
  p_1\shuffle{p_2'}{p_2}p_3$ in normal form.

  The initial state of the DFA is $\iota=(0,0,0,0)$ such that
  $u_\iota=\varepsilon$. The state $p=(i,j,k,\ell)$ is accepting if
  $u_p\equiv w$.

  Our aim is to define the transitions of the automaton in such a way
  that, after reading $v\in\Sigma^*$, the automaton reaches a state
  $p$ with $u_p=\nf{v}$, provided that such a state exists. Furthermore,
  we want to make sure that such a state exists whenever $[v]$ is a
  left-divisor of~$[w]$.

  So let $p=(i,j,k,\ell)$ be a state and $a\in A$. To define the state
  reached from $p$ after reading $a$, let $\ell'>\ell$ be the minimal
  write-position in $w$ after $\ell$. In other words, $\ell<\ell'$,
  $a_{\ell'}\in A$ and $w[\ell+1,\ell'-1]\in{\bar A}^*$. If there is
  no such $\ell'$ or if $a_{\ell'}\neq a$, then the DFA cannot make
  any $a$-move from state $p$. Otherwise, it moves to
  $q=(i,j,k,\ell')$. It is easily verified that this tuple is a state
  again since $p$ is a state and since $a_{\ell'}=a\in A$. We have
  \begin{align*}
    u_p a & = \bar{p_1} \shuffle{p_2'}{p_2} p_3\,a \\
     &= \bar{\bar\pi(w[1,i])} 
        \shuffle{\pi(w[1,k])}{\bar\pi(w[i+1,j])}
        \pi(w[k+1,\ell])\,a\\
     &= \bar{\bar\pi(w[1,i])} 
        \shuffle{\pi(w[1,k])}{\bar\pi(w[i+1,j])}
        \pi(w[k+1,\ell'])\\
     &= u_q\,.
  \end{align*}
      
  \noindent
  We next define which state is reached from $p$ after reading
  $\bar{a}$. Let $j'$ be the minimal read-position in $w$ after
  $j$. In other words, $j<j'$, $a_{j'}\in\bar A$, and $w[j+1,j'-1]\in
  A^*$. If no such $j'$ exists or if $a_{j'}\neq \bar a$, then the DFA
  cannot make any $\bar a$-move from state $p$. So assume $j'$ exists
  with $a_{j'}=\bar a$. Then consider the word
  \[
     s= \OL{\bar\pi(w[i+1,j'])}{\pi(w[1,\ell])}
  \]
  which equals $\OL{p_2a}{p_2p_3}$ since
  $\bar\pi(w[i+1,j'])=\bar\pi(w[i+1,j])a=p_2a$. Since $s$ is a suffix
  of $\bar\pi(w[i+1,j'])$, there exists $i\le i'\le j'$ with
  $s=\bar\pi(w[i'+1,j'])$. In addition, we can assume $i'=0$ or
  $a_{i'}\in\bar{A}$. Similarly, since $s$ is a prefix of
  $\pi(w[1,\ell])$, there exists $1\le k'\le k$ with $s=\pi(w[1,k'])$
  and $k'=0$ or $a_{k'}\in A$. Now the tuple $q=(i',j',k',\ell)$ is a
  state of the DFA and the DFA moves from $p$ to $q$ when reading
  $\bar a$.

  Set
  \begin{align*}
  	r&=p_2a s^{-1}= \bar\pi(w[i+1,j']) \bar\pi(w[i'+1,j'])^{-1}=\bar\pi(w[i+1,i'])
	\quad\text{and} \\
    t&= s^{-1} p_2p_3 = \pi(w[1,k'))^{-1}\pi(w[1,\ell])=\pi(w[k'+1,\ell)) \,.
  \end{align*}
  Then we get
  \begin{align*}
    u_p \bar a 
    &= \bar{p_1}\shuffle{p_2}{p_2} p_3\cdot \bar{a}\\
    \tag{by Theorem~\ref{T-multiplication}}
    &\equiv \bar{p_1} \bar{r} \shuffle{s}{s} t \\
    &= \bar{\bar\pi(w[1,i'])} 
          \shuffle{\pi(w[1,k'])}{\bar\pi(w[i'+1,j'])} \pi(w[k'+1,\ell])\\
    &= u_q\,.
  \end{align*}
  This finishes the construction of the DFA.

  Now let $v\in\Sigma^*$. If there is a $v$-labeled path from the
  initial state $(0,0,0,0)$ to some state $q$, then by induction on
  $|v|$, we obtain $v\equiv u_q$ from the above calculations. In
  particular, any word $v$ accepted by the DFA satisfies $v\equiv w$,
  i.e., $v\in[w]$.

  Before proving the converse implication, let $v\in\Sigma^*$ such
  that $[v]$ is a left-divisor of $[w]$. Let
  $\nf{v}=\bar{v_1}\shuffle{v_2}{v_2}v_3$. Since
  $\bar\pi,\pi\colon\cQ\to A^*$ are morphisms, $v_1v_2$ is a
  prefix of $\bar\pi(w)$ and $v_2v_3$ is a prefix of $\pi(w)$. Hence
  there is a unique state $p=(i,j,k,\ell)$ with $u_p=\nf{v}$. Then, by
  induction on $|v|$, one obtains that there is a $v$-labeled path
  from $(0,0,0,0)$ to $p$. Consequently, for $v\in[w]$, there is a
  $v$\nobreakdash-labeled path from $(0,0,0,0)$ to an accepting state, i.e., the
  DFA accepts~$[w]$.
  
  By the construction of the DFA, it is clear that a Turing machine with $w$ on
  its input tape can, using logarithmic space on its work tape, write the list
  of all transitions on its one-way output tape.
  \qed
\end{proof}

\begin{theorem}
  The following rational subset membership problem for $\cQ$ is
  $\NL$-complete:
  \begin{itemize}
  \item[Input:] a word $w\in\Sigma^*$ and an NFA $\cA$ over $\Sigma$.
  \item[Question:] Is there a word $v\in L(\cA)$ with $w\equiv v$?
  \end{itemize}
\end{theorem}

\begin{proof}
  Let $w\in\Sigma^*$ and let $\cA$ be an NFA over $\Sigma$. Let $\cB$
  be the DFA from Lemma~\ref{L-Automat-fuer-AeK} that can be
  construced in logarithmic space.

  Then there exists $v\in L(\cA)$ with $w\equiv v$ if and only if
  $L(\cA)\cap[w]\neq\emptyset$ if and only if $L(\cA)\cap
  L(\cB)\neq\emptyset$. Using an on-the-fly construction of $\cB$,
  this can be decided nondeterministically in logarithmic space. Hence,
  the problem is in $\NL$.

  Since the free monoid $A^*$ embeds into $\cQ$ and since the rational
  subset membership problem for $A^*$ is $\NL$-hard, we also get
  $\NL$-hardness for $\cQ$.\qed
\end{proof}

\noindent
In the rest of this section, we will prove some negative results on
rational subsets of $\cQ$. All these results rest on a particular
embedding of the monoid ${\{a,b\}^*\times\{c,d\}^*}$ into $\cQ$. This
embedding is discussed in following proposition.

\begin{proposition}\label{P-submonoid}
  Let $\cR = \{[a],[ab],[\bar b],[\bar{abb}]\}^* \subseteq \cQ$
  denote the submonoid generated by
  $\{[a],[ab],[\bar b],[\bar{abb}]\}$.
  \begin{enumerate}[(1)]
  \item There exists an isomorphism $\alpha$ from
    $\{a,b\}^*\times\{c,d\}^*$ onto $\cR$ with
    ${\alpha((a,\varepsilon))=[a]}$, $\alpha((b,\varepsilon))=[ab]$,
    $\alpha((\varepsilon,c))=[\bar b]$, and
    $\alpha((\varepsilon,d))=[\bar{abb}]$.
  \item\label{enum:rec_extends}
    If $\cS\subseteq\cR$ is recognizable in $\cR$, then it is
    recognizable in $\cQ$.
  \end{enumerate}
\end{proposition}

\begin{proof}
  Let $\beta:\{a,b,c,d\}^*\to \cR$ be the morphism defined by
  ${\beta(a)=[a]}$, ${\beta(b)=[ab]}$, $\beta(c)=[\bar b]$, and
  $\beta(d)=[\bar{abb}]$. Note that $\beta$ is surjective.

  Furthermore, note that 
  \[
    \{a,b\}^*\times\{c,d\}^*\cong
    \{a,b,c,d\}^*\mathord{/}\{ac=ca,bc=cb,ad=da,bd=db\}\,.
  \]
  Theorem~\ref{T-multiplication} implies in particular 
  \begin{align*}
    \beta(ac)& =[a\bar b]=[\bar ba]=\beta(ca)\,,\\
    \beta(bc)& =[ab\bar b]=[\bar b ab]=\beta(cb)\,,\\
    \beta(ad)& =[a\bar{abb}]=[\bar{abb} a]=\beta(da)\,,\text{ and}\\
    \beta(bd)& =[ab\bar{abb}]=[\bar{abb} ab]=\beta(db)
  \end{align*}
  since $\OL{\beta(x)}{\beta(y)}=\varepsilon$ for all
  $(x,y)\in\{a,b\}\times\{c,d\}$.

  Hence we can lift $\beta$ to a morphism
  $\alpha\colon\{a,b\}^*\times\{c,d\}^*\to\cR$. The surjectivity of
  $\alpha$ follows from that of $\beta$.

  Note that $\alpha$ maps $\{a,b\}^*\times\{\varepsilon\}$ and
  $\{\varepsilon\}\times\{c,d\}^*$ injectively to disjoint subsets of
  $\cR$. Consequently, $\alpha$ is injective on
  $\{a,b\}^*\times\{c,d\}^*$, i.e., $\alpha$ is an isomorphism as
  required.

  Finally let $\cS\subseteq\cR$ be a recognizable subset of
  $\cR$. Then the subset ${\alpha^{-1}(\cS)\subseteq\{a,b\}^*\times\{c,d\}^*}$ is
  recognizable. By Mezei's theorem, there exist regular languages
  $U_i\subseteq\{a,b\}^*$ and $V_i\subseteq\{c,d\}^*$ with
  $\alpha^{-1}(\cS)=\bigcup_{1\le i\le n} U_i\times V_i$. Define the
  morphism $g\colon\{a,b\}^*\to A^*$ with $g(a)=a$ and $g(b)=ab$
  as well as the morphism $h\colon\{c,d\}^*\to A^*$ with $h(c)=b$
  and $h(d)=abb$. Since morphisms between free monoids preserve
  regularity, the languages $g(U_i),h(V_i)\subseteq A^*$ are
  regular. Therefore, $\pi^{-1}(g(U_i))$ and
  $\bar\pi^{-1}(h(V_i))$ are recognizable in $\cQ$. Hence also
  \[
     \bigcup_{1\le i\le n} \pi^{-1}(g(U_i))\cap\bar\pi^{-1}(h(U_i))
  \]
  is recognizable in $\cQ$. But this set equals $\cS$.\qed
\end{proof}

\begin{theorem}\label{T-rational}
  \begin{enumerate}[(1)]
  \item The set of rational subsets of $\cQ$ is not closed under
    intersection.
  \item The emptiness of the intersection of two rational subsets of
    $\cQ$ is undecidable.
  \item The universality of a rational subset of $\cQ$ is undecidable.

    Consequently, inclusion and equality of rational subsets are undecidable.
  \item The recognizability of a rational subset of $\cQ$ is
    undecidable.
  \end{enumerate}
\end{theorem}

\begin{proof}
  Throughout this proof, let $\alpha$ be the isomorphism from
  Prop.~\ref{P-submonoid}.
  \begin{enumerate}[(1)]
  \item Consider the rational relations
    \[
      R_1=\{(a^m,c^m d^n)\mid m,n\ge1\}\text{ and }
      R_2=\{(a^m,c^n d^m)\mid m,n\ge1\}\,.
    \]
    Then the sets 
    \begin{align*}
      \alpha(R_1) & = \{x\in\cQ\mid 
        \exists m,n\ge1\colon \pi(w)=a^m,\bar\pi(w)=b^m(abb)^n\}
       \quad\text{and}\\
      \alpha(R_2) & = \{x\in\cQ\mid 
        \exists m,n\ge1\colon \pi(w)=a^m,\bar\pi(w)=b^n(abb)^m\}
    \end{align*}
    are rational in $\cQ$. Suppose their intersection
    $\alpha(R_1)\cap\alpha(R_2)$ is rational. Then there exists a
    regular language $S\subseteq\Sigma^*$ with
    \[
      \alpha(R_1)\cap\alpha(R_2) = \nat(S) \,.
    \]
    It follows that the language $\bar\pi(S) \subseteq A^*$
    is regular. But this set equals the language $\{b^m(abb)^m\mid
    m\ge1\}\subseteq\Sigma^*$ which is not regular.

  \item Let $R_1,R_2\subseteq\{a,b\}^*\times\{c,d\}^*$ be
    rational. Then $\alpha(R_1)$ and $\alpha(R_2)$ are rational and,
    since $\alpha$ is an isomorphism,
    $\alpha(R_1)\cap\alpha(R_2)=\alpha(R_1\cap R_2)$. Consequently,
    $\alpha(R_1)\cap\alpha(R_2)=\emptyset$ if and only if $R_1\cap
    R_2=\emptyset$. But this latter question is undecidable
    \cite[Theorem~8.4(i)]{Ber79}.

  \item Let $S\subseteq\{a,b\}^*\times\{c,d\}^*$ be rational. Then
    $\alpha(S)$ is rational.
    Due to Prop.~\ref{P-submonoid}~(\ref{enum:rec_extends}),
    the set $\cR$ is recognizable in $\cQ$.
    Therefore, $\cQ \setminus \cR$ is recognizable and hence rational
    because $\cQ$ is finitely generated.
    Consequently, $\alpha(S)\cup(\cQ\setminus\cR)$ is rational as well.
    This rational set
    equals $\cQ$ if and only if $\alpha(S)=\cR$, i.e.,
    $S=\{a,b\}^*\times\{c,d\}^*$. But this latter question is
    undecidable by \cite[Theorem~8.4(iv)]{Ber79}.

  \item Let $S\subseteq\{a,b\}^*\times\{c,d\}^*$ be rational. Then
    $\alpha(S)$ is rational. By Prop.~\ref{P-submonoid}, $\alpha(S)$
    is recognizable in $\cQ$ if and only if it is recognizably in
    $\cR$. But this is the case if and only if $S$ is recognizable in
    $\{a,b\}^*\times\{c,d\}^*$. This latter question is undecidable by
    \cite[Theorem~8.4(vi)]{Ber79}.\qed
  \end{enumerate}
\end{proof}

\section{Recognizable subsets}
\label{S-recognizable}

\newcommand{\OLset}[1]{\Omega_{#1}}

In this section, we aim to describe the recognizable subsets of $\cQ$.
Clearly,  sets of the form $\pi^{-1}(L)$ or $\bar{\pi}^{-1}(L)$ for some
regular $L \subseteq A^*$ as well as Boolean combinations thereof are
recognizable.  Since definitions of this kind can make no reference to the
relative position of write and read symbols,
there are recognizable sets eluding this form.
For instance, the singleton set $\{ [\bar{a} a] \}$ is recognizable
but any Boolean combination of inverse projections containing $[\bar{a} a]$
also includes $[a \bar{a}]$.
However, we will see in 
the main result of this section, namely Theorem~\ref{thm:recognizability},
that incorporating certain sets that can impose a simple restriction on
these relative positions suffices to generate the recognizable sets as a
Boolean algebra.

Recall Observation~\ref{obs:complete_description}, which states that any
$q \in \cQ$ is completely determined by $\pi(q)$, $\bar{\pi}(q)$, and $\ow{q}$.
Consequently, it would seem natural to incorporate sets which restrict the overlap width.
Unfortunately, the overlap width is not a recognizable property in the following sense:

\begin{observation}
\label{obs:ow_non_rec}
For each $k \in \bN$, the set of all $q \in \cQ$ with $\ow{q} = k$ is not recognizable.
\end{observation}

\begin{proof}
It suffices to show that the set
\begin{equation*}
	L_k = \set{ w \in \Sigma^* | \ow{w} = k }
\end{equation*}
is not regular. For the sake of a contradiction, suppose there was a finite automaton $\cA$
recognizing $L_k$. Let $n \geq k$ be an upper bound on the number of states of $\cA$.
Consider the word $w = a^n b a^k \bar{a}^{n-1} \bar{b} \bar{a}^k$.
Since $\nf{w} = \bar{a}^{n-1} \bar{b} \eshuffle{a^k}{\bar{a}^k} a^{n-k} b a^k$, we have
$\ow{w} = k$, i.e., $w \in L_k$. Therefore, $\cA$ accepts $w$.
Using a pumping argument, we obtain $\ell \leq n-1$ such that $\cA$ also accepts
$w' = a^\ell b a^k \bar{a}^{n-1} \bar{b} \bar{a}^k$.
However, $\nf{w'} = \bar{a}^{n-1-\ell} \eshuffle{a^\ell b a^k}{\bar{a}^\ell \bar{b} \bar{a}^k}$
implies $\ow{w} = \ell+1+k > k$ and hence $w \not\in L_k$.
Contradiction.
\qed
\end{proof}

\noindent
In fact, the proof above also shows that the set of all $q \in \cQ$ with $\ow{q} \leq k$
is not recognizable for any $k \in \bN$. Thus, the set of all
$q \in \cQ$ with $\ow{q} > k$ is not recognizable either.

Nevertheless, the definition below provides a slight variation of this idea
conducing to our purpose.
To simplify notation, we say two elements ${p,q \in \cQ}$ \emph{have the same projections}
and write $p \sim_\pi q$ if $\pi(p) = \pi(q)$ and $\bar{\pi}(p) = \bar{\pi}(q)$.

\begin{definition}
\label{def:olset_k}
For each $k \in \bN$, the set $\OLset{k} \subseteq \cQ$ is given by
\begin{equation*}
	\OLset{k} = \set{ q \in \cQ | \forall p \in \cQ \colon
		p \sim_\pi q \mathbin{\&} \ow{q} \leq \ow{p} \leq k \implies p = q } \,.
\end{equation*}
\end{definition}

\noindent
Observe that $\cQ = \OLset{0} \supseteq \OLset{1} \supseteq \OLset{2} \supseteq \dotsc$.
Intuitively, for fixed projections $\pi(q)$ and $\bar{\pi}(q)$ the set $\OLset{k}$ contains
all $q$ with $\ow{q} \geq k$ as well as the unique $q$ with maximal $\ow{q} \leq k$.
From this perspective, the set $\OLset{k}$ is similar to the set in Observation~\ref{obs:ow_non_rec}
but uses an overestimation of the overlap width instead of the overlap width itself.

\begin{example}
\begin{enumerate}[(1)]
\item The queue action $q = [\bar{a} \bar{b} a \bar{a} b a]$ satisfies $\ow{q} = 1$ and
hence $q \in \OLset{1}$. The only $p \in \cQ$ with $p \sim_\pi q$ and $\ow{p} \geq \ow{q}$
is $p = [a \bar{a} b \bar{b} a \bar{a}]$. Since $\ow{p} = 3$, this implies
$q \in \OLset{2}$ but $q \not\in \OLset{3}$.
\item For every $k \geq 1$, we have $[(\bar{a} a)^k] \in \OLset{k-1} \setminus \OLset{k}$.
\item All queue actions of the form $q = [u \bar{v}]$ with
$u,v \in A^*$ satisfy $q \in \OLset{k}$ for every $k \in \bN$.
\end{enumerate}
\end{example}

\noindent
The following observation is to the sets $\OLset{k}$ as
Observation~\ref{obs:complete_description} is to the overlap width
and provides some more motivation for defining the sets $\OLset{k}$.

\begin{observation}
Every $q \in \cQ$ is completely described by $\pi(q)$, $\bar{\pi}(q)$,
and the maximal $k \in \bN$ with $q \in \OLset{k}$ or the fact that there is no such maximum.
\end{observation}

\begin{proof}
Fix $u,v \in A^*$ and consider some $q \in \cQ$ with $\pi(q) = u$ and $\bar{\pi}(q) = v$.
Let $m = \max \set{ k \in \bN | q \in \OLset{k} }$ or $m = \infty$ if this
maximum does not exist.
Due to Observation~\ref{obs:complete_description},
it suffices to provide $\ow{q}$ in terms of $u$, $v$, and $m$.
To this end, let $w \in A^*$ be the longest suffix of $v$
that is also a prefix of $u$ and satisfies $|w| \leq m$.
In particular, we have $q \in \OLset{|w|}$.
We claim that $\ow{q} = |w|$.

First, we have $\ow{q} \leq m$. This is trivial for $m = \infty$ and follows directly
from $q \not\in \OLset{m+1}$ for $m < \infty$.
Since there is a suffix of length $\ow{q}$ of $\bar{\pi}(q) = v$ that is also
a prefix of $\pi(q) = u$ and due to the maximality of the length of $w$,
we may conclude $\ow{q} \leq |w|$.
The choice of $w$ further implies the existence of some $p \in \cQ$ with
$p \sim_\pi q$ and $\ow{p} = |w|$.
From $q \in \OLset{|w|}$ and $\ow{q} \leq \ow{p} \leq |w|$,
we conclude $p = q$ and hence $\ow{q} = |w|$.
\qed
\end{proof}

\noindent
The aforementioned main result of this section characterizing the recognizable subsets of $\cQ$
is Theorem~\ref{thm:recognizability} below.

\begin{theorem}
\label{thm:recognizability}
For every subset $L \subseteq \cQ$, the following are equivalent:
\begin{enumerate}[(1)]
\item\label{enum:recognizable}
$L$ is recognizable,
\item\label{enum:wrw-recognizable}
 $\nat^{-1}(L) \cap A^* \bar{A}^* A^*$ is regular,
\item\label{enum:rwr-recognizable}
$\nat^{-1}(L) \cap \bar{A}^* A^* \bar{A}^*$ is regular,
\item\label{enum:simple}
$L$ is a Boolean combination of sets of the form
$\pi^{-1}(R)$ or $\bar{\pi}^{-1}(R)$ for some regular $R \subseteq A^*$ and
the sets $\OLset{k}$ for $k \in \bN$.
\end{enumerate}
\end{theorem}

\noindent
The implication \describeImp{enum:recognizable}{enum:wrw-recognizable} is trivial.
Throughout the rest of this section, we call subsets $L \subseteq \cQ$ satisfying
condition~(\ref{enum:wrw-recognizable}) above \emph{wrw-recognizable}.
The motivation behind wrw-recognizability is a follows:
Consider a queue action $q \in \cQ$ and let $\nf{q} = \bar{u} \shuffle{v}{v} w$.
Lemma~\ref{lemma:overlap} and Corollary~\ref{cor:contextcommute} yield
$\bar{u} \shuffle{v}{v} w \equiv \bar{u} v \bar{v} w \equiv v \bar{u v} w$, i.e.,
$q = [v \bar{u v} w]$.
Thus, we have $q \in L$ if and only if
$\nat^{-1}(L) \cap A^* \bar{A}^* A^*$ contains at least one representative of $q$,
although it might include even more than one representative.
Finally, notice that condition~(\ref{enum:rwr-recognizable}) is dual to
condition~(\ref{enum:wrw-recognizable}).

A complete proof of Theorem~\ref{thm:recognizability} follows at the end of this section.
Our first step into this direction is
to demonstrate the implication \describeImp{enum:simple}{enum:recognizable}.
Basically, we only have to show that $\OLset{k}$ is recognizable for each $k \in \bN$
(see Proposition~\ref{prop:OLset_recognizable}).
To this end, we say that a word $w \in \Sigma^*$ is \emph{$k$-shuffled} if
it contains at least $k$ write and $k$ read symbols, respectively, and for each
$i = 1,\dotsc,k$ the $i$-th write symbol of $w$ appears before the $i$-th of the
last $k$ read symbols of $w$.
We need the following relationship between the overlap width and $k$-shuffledness.

\begin{lemma}
\label{lemma:olwidth_vs_shuffledness}
Let $k \in \bN$, $w \in \Sigma^*$, and
$u \in A^k$ a prefix of $\pi(w)$ as well as a suffix of $\bar{\pi}(w)$.
Then $w$ is $k$-shuffled if and only if $\ow{w} \geq k$.
\end{lemma}

\begin{proof}
We show both claims by induction on $n \in \bN$ with $w \xrightarrow{n} \nf{w}$.
If $n = 0$, then $w$ is in normal form and the claim is obvious.

Henceforth, we assume $n > 0$. Let $w' \in \Sigma^*$ with
$w \to w' \xrightarrow{n-1} \nf{w}$.
In particular, there are $x,y \in \Sigma^*$ and $a,b \in A$ such that
$w = x a \bar{b} y$ and ${w' = x \bar{b} a y}$.
By the induction hypothesis, the claim holds for $w'$.
As we have $\pi(w) = \pi(w')$, ${\bar{\pi}(w) = \bar{\pi}(w')}$, and $\ow{w} = \ow{w'}$,
it suffices to show that $w$ is $k$-shuffled if and only if $w'$ is $k$-shuffled.
The ``if''-part is easy to check even without using $u$.

The claim of the ``only if''-part is trivial unless
$a$ is among the first $k$ write symbols of $w$, say the $i$-th of them,
and $\bar{b}$ among the last $k$ read symbols of $w$, say the $j$-th of them.
If $i > j$, then the $i$-th of the last $k$ read symbols of $w$ is contained in $y$
and the $j$-th write symbol of $w$ is contained in $x$. Thus, $w'$ is also $k$-shuffled.
We cannot have $i < j$, because then the $j$-th write symbol of $w$ would have to appear
after $a$ but before $\bar{b}$.

Finally, we show that $i = j$ is also impossible. According to the exact rule used
in $w \to w'$, we distinguish three cases. If $a = b$ and the rule was
$c a \bar{b} \to c \bar{b} a$ for some $c \in A$, then $i > 1$ and the $(i-1)$-th
of the last $k$ read symbols of $w$ would have to appear after $x$ but before $\bar{b}$.
Dually, if $a = b$ and the rule was $a \bar{b} \bar{c} \to \bar{b} a \bar{c}$ for some $c \in A$,
then $j < k$ and the $(j+1)$-th write symbol would have to appear after $a$ but before $y$.
If $a \not= b$ and the rule was $a \bar{b} \to \bar{b} a$, this would contradict the fact
the $i$-th write symbol of $w$ as well as the $i$-th of the last $k$ read symbols of $w$
coincide with the $i$-th symbol of $u$.
\qed
\end{proof}

\begin{lemma}
\label{lemma:olset_k_characterization}
For each $k \in \bN$, we have
\begin{equation*}
	\nat^{-1}(\OLset{k})
	= \Set{ w \in \Sigma^* | \forall u \in A^{\leq k}\colon
	    \genfrac{}{}{0pt}{0}{\text{$u$ prefixes $\pi(w)$} \mathrel{\&}}{\text{$u$ suffixes $\bar{\pi}(w)$}}
		\implies \text{$w$ is $|u|$-shuffled} } \,.
\end{equation*}
\end{lemma}

\begin{proof}
Denote the set on the right hand side by $Z_k$.
First, suppose $w \in \nat^{-1}(\OLset{k})$ and
consider some $u \in A^{\leq k}$ that is a prefix of $\pi(w)$ as well as a suffix of $\bar{\pi}(w)$.
Let $x,y \in A^*$ such that $\pi(w) = uy$ and $\bar{\pi}(w) = xu$. The queue action
$p = [\bar{x} \shuffle{u}{u} y]$ satisfies $p \sim_\pi [w]$ and
 $\ow{p} = |u| \leq k$. Since $[w] \in \OLset{k}$,
this implies $|u| = \ow{p} \leq \ow{w}$. By Lemma~\ref{lemma:olwidth_vs_shuffledness},
we obtain that $w$ is $|u|$-shuffled and hence $w \in Z_k$.

Now, assume $w \in Z_k$ and consider some $p \in \cQ$ with $p \sim_\pi [w]$ and
$\ow{w}\le\ow{p} \leq k$. Let $\nf{p} = \bar{x} \shuffle{u}{u} y$.
Then $|u| = \ow{p} \leq k$ and $u$ is a prefix of $\pi(p) = \pi(w)$ as well as a suffix
of $\bar{\pi}(p) = \bar{\pi}(w)$. Since $w \in Z_k$, this implies that $w$ is
$|u|$\nobreakdash-shuffled. From Lemma~\ref{lemma:olwidth_vs_shuffledness}, we finally conclude
$\ow{w} \geq |u| = \ow{p}$. This proves $[w] \in \OLset{k}$, i.e.,
$w \in \nat^{-1}(\OLset{k})$.
\qed
\end{proof}

\begin{proposition}
\label{prop:OLset_recognizable}
For each $k \in \bN$, the set $\OLset{k}$ is recognizable.
\end{proposition}

\begin{proof}
It suffices to show
that the set $\nat^{-1}(\OLset{k})$ is regular. For $\ell \in \bN$, let $S_\ell$
denote the set of all $w \in \Sigma^*$ that are $\ell$-shuffled.
Lemma~\ref{lemma:olset_k_characterization} translates directly into
\begin{equation*}
	\nat^{-1}(\OLset{k})
	= \bigcap_{u \in A^{\leq k}}
		\Sigma^* \setminus \left( \pi^{-1}(uA^*) \cap \bar{\pi}^{-1}(A^*u) \right)
		\cup S_{|u|} \,.
\end{equation*}
Thus, it only remains to show that all the sets $S_\ell$ for $\ell \leq k$ are
regular. A word $w \in \Sigma^*$ is $\ell$-shuffled if and only if it admits
for each $i = 1,\dotsc,\ell$ a factorization $w = x_i a_i y_i \bar{b_i} z_i$ with
$x_i,y_i,z_i \in \Sigma^*$, $a_i,b_i \in A$, $|\pi(x_i)| = i-1$, and
$|\bar{\pi}(z_i)| = \ell-i$ ($a_i$ is the $i$-th write symbol,
$\bar{b_i}$ the $i$-th of the last $\ell$ read symbols). This translates directly
into
\begin{equation*}
	\tag*{\qed}
	S_\ell
	= \bigcap_{1 \leq i \leq \ell}
		\pi^{-1}(A^{i-1}) \, A \, \Sigma^* \, \bar{A} \, \bar{\pi}^{-1}(A^{\ell-i}) \,.
\end{equation*}
\end{proof}

\noindent
Our next step towards proving Theorem~\ref{thm:recognizability} is to establish the
implication \describeImp{enum:wrw-recognizable}{enum:simple}
(see Proposition~\ref{prop:wrw-rec_implies_simple}).
Again, we prepare this by a series of lemmas.
Throughout, we call a subset $L \subseteq \cQ$ \emph{simple} if is satisfies
condition~(\ref{enum:simple}) of Theorem~\ref{thm:recognizability}.
Recall that sets meeting condition~(\ref{enum:wrw-recognizable}) are called
\emph{wrw-recognizable}.

\begin{lemma}
\label{lemma:prefixes}
Let $k \in \bN$, $q \in \OLset{k}$, and $u \in A^k$ be a prefix of $\pi(q)$.
Then there exists $p \in \cQ$ such that $q = [u] \, p$.
\end{lemma}

\begin{proof}
Let $\nf{q} = \bar{x} \shuffle{y}{y} z$. If $u$ is already a prefix of $y$, say $y = uv$,
we choose $p = [v \bar{x} \bar{y} z]$ and obtain
$q = [u v \bar{x} \bar{y} z] = [u] \, p$.
Now, suppose that $u$ is not a prefix of $y$. Then there is a prefix $v$ of $z$,
say $z = vw$, such that $u = yv$. The queue action $r = [yv \bar{x} \bar{y} w]$
satisfies $r \sim_\pi q$ and $\ow{r} \leq |yv| = k$. Since $q \in \OLset{k}$,
this implies $\ow{r} \leq \ow{q} = |y|$. At the same time, $\ow{r} \geq |y|$ and
hence $q = r$.
Thus, we obtain $q = [u] \, p$ for $p = [\bar{x} \bar{y} w]$.
\qed
\end{proof}

\begin{lemma}
\label{lemma:boolean_decomp_1}
Let $k \in \bN$ and $L \subseteq \cQ$. If $L$ is wrw-recognizable,
then the following set is simple:
\begin{equation*}
	L \cap \pi^{-1}\left(A^{< k}\right) \cap \OLset{k} \,.
\end{equation*}
\end{lemma}

\begin{proof}
Let $K = \nat^{-1}(L) \cap A^* \bar{A}^* A^*$ and
$\phi\colon \Sigma^* \to M$ be a morphism recognizing $K$.
We further consider the morphisms
$\mu,\bar{\mu}\colon A^* \to M$ defined by $\mu(w) = \phi(w)$ and
$\bar{\mu}(w) = \phi(\bar{w})$.
We show the claim by establishing the equation
\begin{equation*}
	L \cap \pi^{-1}\left(A^{<k}\right) \cap \OLset{k}
	= \bigcup_{\substack{u \in A^{<k}, m \in M \\ \mu(u) m \in \phi(K)}}
		\pi^{-1}(u) \cap \bar{\pi}^{-1}\left(\bar{\mu}^{-1}(m)\right) \cap \OLset{k} \,.
\end{equation*}
Let $X$ and $Y$ denote the left and right hand side of this equation, respectively.
Clearly, ${X, Y \subseteq \pi^{-1}\left(A^{< k}\right) \cap \OLset{k}}$.
Consider some $q \in \pi^{-1}\left(A^{< k}\right) \cap \OLset{k}$.
It suffices to show that $q \in X$ precisely if $q \in Y$.

To this end, let $u = \pi(q)$. Then $|u| < k$ and hence $q \in \OLset{k}\subseteq\OLset{|u|}$.
Due to Lemma~\ref{lemma:prefixes}, there is $p \in \cQ$ such that $q = [u] \, p$.
Clearly, $\pi(p) = \varepsilon$, i.e., $p = [\bar{y}]$ for some $y \in A^*$.
Notice that $q = [u \bar{y}]$. Altogether,
\begin{align*}
	q \in X
	&\quad\Longleftrightarrow\quad
	q = [u \bar{y}] \in L \\
	&\quad\Longleftrightarrow\quad
	\phi(u \bar{y})
	= \mu\left(u\right) \, \bar{\mu}\left(\bar{\pi}(q)\right) \in \phi(K)
	\quad\Longleftrightarrow\quad
	q \in Y \,.
	\tag*{\qed}
\end{align*}
\end{proof}

\begin{lemma}
\label{lemma:boolean_decomp_2}
Let $k \in \bN$ and $L \subseteq \cQ$. If $L$ is wrw-recognizable
by a monoid with $k$ elements, then the following set is simple:
\begin{equation*}
	L \cap \pi^{-1}\left(A^{\geq k}\right) \cap \OLset{k} \,.
\end{equation*}
\end{lemma}

\begin{proof}
Let $K$, $\phi$, $M$, $\mu$, and $\bar{\mu}$ be as in the proof of
Lemma~\ref{lemma:boolean_decomp_1} and additionally assume that $|M| = k$.
We show the claim by establishing the equation
\begin{equation*}
\label{eq:boole_comb_1}
	L \cap \pi^{-1}\left(A^{\geq k}\right) \cap \OLset{k}
	= \bigcup_{\substack{u \in A^k, m,m' \in M\\\mu(u) m' m \in \phi(K)}}
		\pi^{-1}\left(u\mu^{-1}(m)\right) \cap
		\bar{\pi}^{-1}\left(\bar{\mu}^{-1}(m')\right) \cap
		\OLset{k} \,.
\end{equation*}
Once more, call the left and right hand side $X$ and $Y$, respectively.
Clearly, ${X, Y \subseteq \pi^{-1}\left(A^{\geq k}\right) \cap \OLset{k}}$.
Consider some $q \in \pi^{-1}\left(A^{\geq k}\right) \cap \OLset{k}$.
It suffices to show that $q \in X$ precisely if $q \in Y$.

Since $|\pi(q)| \geq k$, there is a prefix $u \in A^k$ of $\pi(q)$.
Lemma~\ref{lemma:prefixes} provides us with $p \in \cQ$ satisfying $q = [u] \, p$.
According to the motivation of wrw-recognizability right below
Theorem~\ref{thm:recognizability}, there are $x,y,z \in A^*$ with $p = [x \bar{y} z]$.
Notice that $q = [u x \bar{y} z]$.
Since $|M| = k$, there is $y_0 \in A^{\leq k}$ such that $\phi(\bar{y_0}) = \phi(\bar{y})$.
Due to $|u| = k \geq |y_0|$ and Corollary~\ref{cor:contextcommute},
we conclude $u x \bar{y_0} z \equiv u \bar{y_0} x z$. Combining these facts yields
\begin{align*}
	q \in L
	&\quad\Longleftrightarrow\quad
	\phi(u x \bar{y} z) \in \phi(K) 
		&\qquad& \text{since $q = [u x \bar{y} z]$} \\
	&\quad\Longleftrightarrow\quad
	\phi(u x \bar{y_0} z) \in \phi(K)
		&& \text{since $\phi(u x \bar{y} z) = \phi(u x \bar{y_0} z)$} \\
	&\quad\Longleftrightarrow\quad
	[u x \bar{y_0} z] \in L \\
	&\quad\Longleftrightarrow\quad
	[u \bar{y_0} x z] \in L
		&& \text{since $u x \bar{y_0} z \equiv u \bar{y_0} x z$} \\
	&\quad\Longleftrightarrow\quad
	\phi(u \bar{y_0} x z) \in \phi(K) \\
	&\quad\Longleftrightarrow\quad
	\phi(u \bar{y} x z) \in \phi(K)
		&& \text{since $\phi(u \bar{y_0} x z) = \phi(u \bar{y} x z)$} \,.
\end{align*}
Moreover, we have
\begin{equation*}
	\phi(u \bar{y} x z)
	= \mu\left(u\right) \, \bar{\mu}\left(\bar{\pi}(q)\right) \, \mu\left(u^{-1}\pi(q)\right) \,.
\end{equation*}
As we assumed that $q \in \pi^{-1}(A^{\geq k}) \cap \OLset{k}$, we obtain
\begin{equation*}
	q \in X
	\quad\Longleftrightarrow\quad
	q \in L
	\quad\Longleftrightarrow\quad
	\mu\left(u\right) \,
		\bar{\mu}\left(\bar{\pi}(q)\right) \,
		\mu\left(u^{-1}\pi(q)\right) \in \phi(K) \,.
\end{equation*}
Finally, utilizing $m = \mu\left(u^{-1} \pi(q)\right)$ and
$m' = \bar{\mu}\left(\bar{\pi}(q)\right)$ reveals that the last condition above
is equivalent to $q \in Y$.
\qed
\end{proof}

\begin{lemma}
\label{lemma:boolean_decomp_3}
Let $k \in \bN$ and $L \subseteq \cQ$. If $L$ is wrw-recognizable,
then the following set is simple:
\begin{equation*}
	L \cap \OLset{k} \setminus \OLset{k+1} \,.
\end{equation*}
\end{lemma}

\begin{proof}
Let $K$, $\phi$, $M$, $\mu$, and $\bar{\mu}$ be as in the proof of
Lemma~\ref{lemma:boolean_decomp_1}.
We show the claim by establishing the equation
\begin{equation*}
	L \cap \OLset{k} \setminus \OLset{k+1}
	= \bigcup_{\substack{u \in A^k, m,m' \in M \\
			\mu(u) m' m \in \phi(K)}}
		\pi^{-1}\left(u \mu^{-1}(m)\right) \cap
		\bar{\pi}^{-1}\left(\bar{\mu}^{-1}(m')\right) \cap
		\OLset{k} \setminus \OLset{k+1} \,.
\end{equation*}
Again, call the two sides $X$ and $Y$, respectively.
Clearly, ${X,Y \subseteq \OLset{k} \setminus \OLset{k+1}}$.
Consider some $q \in \OLset{k} \setminus \OLset{k+1}$.
It suffices to show that $q \in X$ precisely if $q \in Y$.

Since $q \not\in \OLset{k+1}$, there is $p_0 \in \cQ$ with $p_0 \sim_\pi q$,
$\ow{p_0} \leq k+1$, and ${\ow{p_0} > \ow{q}}$. As ${\ow{p_0} \leq k}$
would contradict $q \in \OLset{k}$, we have $\ow{p_0} = k+1$
and hence $\ow{q} \leq k$. Thus, there are
$u \in A^k$ and $a \in A$ such that $ua$ is a prefix of $\pi(p_0) = \pi(q)$
and a suffix of $\bar{\pi}(p_0) = \bar{\pi}(q)$.
In particular, $u$ is a prefix of $\pi(q)$ and by Lemma~\ref{lemma:prefixes}
there is $p \in \cQ$ with $q = [u] \, p$. There are $x,y,z \in A^*$ with
$p = [x \bar{y} z]$. Notice that $q = [u x \bar{y} z]$, $a$ is a prefix
of $xz$, and $ua$ is a suffix of $y$. Due to the latter and $\ow{q} \leq k$,
$a$ cannot be a prefix of $x$, i.e., $x = \varepsilon$. Altogether, we obtain
\begin{align*}
	q \in X
	&\quad\Longleftrightarrow\quad
	\phi(u \bar{y} z) \in \phi(K)
		&\quad& \text{since $q = [u \bar{y} z]$} \\
	&\quad\Longleftrightarrow\quad
	q \in Y
		&& \text{since $\phi(u \bar{y} z) = \mu\left(u\right) \,
		\bar{\mu}\left(\bar{\pi}(q)\right)
		\mu\left(u^{-1}\pi(q)\right)$} \,,
\end{align*}
where the last equivalence again uses
$m = \mu\left(u^{-1} \pi(q)\right)$ and
$m' = \bar{\mu}\left(\bar{\pi}(q)\right)$.
\qed
\end{proof}

\begin{proposition}
\label{prop:wrw-rec_implies_simple}
Every wrw-recognizable subset $L \subseteq \cQ$ is simple.
\end{proposition}

\begin{proof}
Suppose that $\nat^{-1}(L) \cap A^* \bar{A}^* A^*$ is recognizable by
a monoid with $k$ elements. Since
$\cQ = \OLset{0} \supseteq \OLset{1} \supseteq \dotsm \supseteq \OLset{k}$,
we have
\begin{equation*}
	L
	= \left( L \cap \pi^{-1}\left(A^{< k}\right) \cap \OLset{k} \right) \cup
		\left( L \cap \pi^{-1}\left(A^{\geq k}\right) \cap \OLset{k} \right) \cup
		\bigcup_{0 \leq \ell < k} \left( L \cap \OLset{\ell} \setminus \OLset{\ell+1} \right)
	\,.
\end{equation*}
By Lemmas~\ref{lemma:boolean_decomp_1}, \ref{lemma:boolean_decomp_2},
and~\ref{lemma:boolean_decomp_3}, the right hand side is a finite union of simple sets
and a simple set itself.
\qed
\end{proof}

\noindent
We are now prepared to prove the main result of this section.

\begin{proof}[of Theorem~\ref{thm:recognizability}]
We establish the circular chain of implications
``(\ref{enum:recognizable})$\Rightarrow$%
(\ref{enum:wrw-recognizable})$\Rightarrow$%
(\ref{enum:simple})$\Rightarrow$(\ref{enum:recognizable})''
as well as the equivalence
\describeEq{enum:recognizable}{enum:rwr-recognizable}.

\medskip\noindent
\emph{To \describeImp{enum:recognizable}{enum:wrw-recognizable} and
\describeImp{enum:recognizable}{enum:rwr-recognizable}.}\ 
Since $L$ is recognizable, $\nat^{-1}(L)$ is regular and
the claims follow.

\medskip\noindent
\emph{To \describeImp{enum:wrw-recognizable}{enum:simple}.}\ 
This is precisely the statement of Proposition~\ref{prop:wrw-rec_implies_simple}.

\medskip\noindent
\emph{To \describeImp{enum:simple}{enum:recognizable}.}\ 
For regular $L \subseteq A^*$, the sets $\pi^{-1}(L)$ and $\bar{\pi}^{-1}(L)$
are recognizable. The sets $\OLset{k}$ with $k \in \bN$ are recognizable
by Proposition~\ref{prop:OLset_recognizable}. Since the class of recognizable subsets
of $\cQ$ is closed under Boolean combinations, the claim follows.

\medskip\noindent
\emph{To \describeImp{enum:rwr-recognizable}{enum:recognizable}.}\ 
Let $K = \nat^{-1}(L) \cap \bar{A}^* A^* \bar{A}^*$. Then
\begin{equation*}
	\dual(K)
	= \dual\left(\nat^{-1}(L)\right) \cap \dual\left(\bar{A}^* A^* \bar{A}^*\right)
	= \nat^{-1}\left(\dual(L)\right) \cap A^* \bar{A}^* A^* \,.
\end{equation*}
Since $K$ is regular, $\dual(K)$ is regular as well and
the already established implication
\describeImp{enum:wrw-recognizable}{enum:recognizable}
yields that $\dual(L)$ is recognizable.
Finally, this implies that $L$ is recognizable.
\qed
\end{proof}

\noindent
In light of Theorem \ref{thm:recognizability}, the question arises whether the
regularity of ${\nat^{-1}(L) \cap \bar{A}^* A^*}$ or of
$\nat^{-1}(L) \cap A^* \bar{A}^*$ or of both of them
already suffices to conclude recognizability of $L$.
The answer is negative, as demonstrated by the following example.
The set $L = \set{ [\bar{a}^na\bar{a}a^n] | n \geq 1 }$ is not recognizable,
since the set of its normal forms is not regular.
However, both of the sets $\nat^{-1}(L) \cap \bar{A}^* A^*$ and
$\nat^{-1}(L) \cap A^* \bar{A}^*$ are empty and hence regular.

\section{Thurston-automaticity}
\label{S-automatic}

Many groups of interest in combinatorial group theory turned out to be
Thurston-automatic~\cite{EPCHLT1992}. The more general concept of a
Thurston-automatic semigroup was introduced in~\cite{CaRoRuTh2001}. In
this chapter, we prove that the monoid of queue-actions $\cQ$ does not
fall into this class. 

Let $\Gamma$ be an alphabet and $\diamond\notin\Gamma$. Then consider
the new alphabet $\Gamma(2,\diamond) =
(\Gamma\cup\{\diamond\})^2\setminus\{(\diamond,\diamond)\}$. We define
the \emph{convolution} $\otimes : \Gamma^* \times \Gamma^* \to
\Gamma(2,\diamond)^*$ as follows:
\begin{align*}
 \varepsilon\otimes\varepsilon &= \varepsilon &
 av\otimes\varepsilon &= (a,\diamond)(v\otimes\varepsilon) &
 \varepsilon\otimes bw &= (\diamond,b)(\varepsilon\otimes w)\\
 &&av\otimes bw &=(a,b)(v\otimes w)
\end{align*}
for $a,b\in\Gamma$ and $v,w\in\Gamma^*$. If
$R\subseteq\Gamma^*\times\Gamma^*$ let
\[
  R^\otimes=\set{ v\otimes w | (v,w)\in R }
\]
denote the convolution of~$R$. Note that $R^\otimes$ is a language over
the alphabet $\Gamma(2,\diamond)$. 

Let $M$ be a monoid, $\Gamma$ an alphabet, $\theta\colon \Gamma^+\to
M$ a semigroup morphism, $L\subseteq\Gamma^+$, and
$a\in\Gamma$. Then we define:
\[
   L_a=\Set{(u,v)\in L^2 | \theta(ua)=\theta(v) }^\otimes\,.
\]
The triple $(\Gamma,\theta,L)$ is an \emph{automatic presentation} for
the monoid $M$ if $\theta$ maps $L$ bijectively onto $M$ and if the
languages $L$ and $L_a$ for all $a\in\Gamma$ are
regular.\footnote{This is not the original definition from
  \cite{CaRoRuTh2001}, but it is equivalent by
  \cite[Prop.~5.4]{CaRoRuTh2001}.} A monoid is \emph{Thurston-automatic}
  if it has some automatic presentation.

Two fundamental results on automatic monoids are the following:

\begin{proposition}\label{P-DanRR99}
  Let $M$ be a Thurston-automatic monoid.
  \begin{enumerate}
  \item If $(\Gamma,\theta,L)$ is an automatic presentation of $M$ and
    $b\in\Gamma$, then the language
    \[
       \set{u\otimes v | u,v\in L, \theta(ub)=\theta(vb) }
    \]
    is regular \cite{CaRoRuTh2001}.
  \item If $\Gamma$ is a finite set and $\mu\colon\Gamma^*\to M$ a
    surjective morphism, then $M$ admits an automatic presentation
    $(\Gamma\cup\{\iota\},\theta,L)$ for some $\iota\notin\Gamma$ with
    $\theta(a)=\mu(a)$ for all $a\in\Gamma$ and $\theta(\iota)=1$
    \cite{DuRoRu1999}.
  \end{enumerate}
\end{proposition}

\noindent
Using only these basic properties of Thurston-automatic monoids (and a
simple counting argument), we can show that $\cQ$ does not admit an
automatic presentation.

\begin{theorem}
  The monoid of queue actions $\cQ$ is not Thurston-automatic.
\end{theorem}

\begin{proof}
  Aiming towards a contradiction, assume $\cQ$ to be
  Thurston-automatic. Recall that, by the very definition, $\cQ$ is
  generated by the set $\Sigma= A\cup\bar{A}$ and hence the natural
  morphism $\nat\colon\Sigma^*\to\cQ$ is surjective. Throughout this
  proof, let $a,b\in A$ be two distinct letters. By
  Prop.~\ref{P-DanRR99}(2), there exists an automatic presentation
  $(\Sigma\cup\{\iota\},\theta,L)$ with $\theta(c)=\nat(c)$ for all
  $c\in\Sigma$ and $\theta(\iota)=\nat(\varepsilon)$. Let
  $\varphi:(\Sigma\cup\{\iota\})^*\to\Sigma^*$ be the morphism with
  $\varphi(c)=c$ for $c\in\Sigma$ and
  $\varphi(\iota)=\varepsilon$. Since $\varphi(\iota)=\varepsilon$ and
  since $\theta$ agrees with $\eta$ on $\Sigma^*$, we get
  $\theta(v)=\theta(\varphi(v))=\nat(\varphi(v))$ for all
  $v\in(\Sigma\cup\{\iota\})^*$.

  By Prop.~\ref{P-DanRR99}(1), the relation
  \[
     R_0 = \{(u,v)\in L^2\mid \theta(u\bar b)=\theta(v\bar b)\}
  \]
  is synchronously rational. Since $\varphi$ is a morphism, also the
  relation
  \[
     R = \{(\varphi(u),\varphi(v))\mid 
         u,v\in L, \theta(u\bar b)=\theta(v\bar b)\}
  \]
  is rational~\cite{Ber79}. For $(\varphi(u),\varphi(v))\in R$, we
  have $\nat(\varphi(u)\bar b)=\theta(u\bar b)=\theta(v\bar
  b)=\nat(\varphi(v)\bar b)$ and therefore
  $|\varphi(u)|=|\varphi(v)|$. It follows that the relation $R$ is
  synchronously rational \cite{FroS93}, i.e., that the language
  $R^\otimes$ is regular.

  Let $m,n\in\bN$. Since $\theta|_L$ maps $L$ bijectively onto $\cQ$,
  there is a unique word $u_{m,n}\in L$ with
  $\theta(u_{m,n})=[\bar{a}^m a^n]$. Then we have
  $\eta(\varphi(u_{m,n}))=\theta(u_{m,n})=[\bar a^m a^n]$. Since $\bar
  a^m a^n$ is the only element of $[\bar a^m a^n]$, this implies
  $\varphi(u_{m,n})=\bar a^m a^n$.

  For $q\in \cQ$, $\theta(u_{m,n})[\bar b]= q[\bar b]$ is equivalent to
  saying $\pi(q)=a^n$ and $\bar\pi(q)=a^m$ (the implication
  ``$\Rightarrow$'' is trivial since $\pi$ and $\bar\pi$ are
  morphisms, the converse one follows from
  Theorem~\ref{T-multiplication}). Since $q\in \cQ$ is determined by the
  projections and the overlap width $\ow{q}$, there are precisely
  $\min(m,n)+1$ many elements $q\in \cQ$ with $\theta(u_{m,n})[\bar
  b]=q[\bar b]$. Since $\theta$ is bijective on $L$, there are
  precisely $\min(m,n)+1$ many words $v\in L$ with $(u_{m,n},v)\in
  R$. Since also $\varphi$ is injective on $L$, we get
  \begin{align*}
    \min(m,n)+1 & = |\{\varphi(v)\mid (u_{m,n},v)\in R_0\}|\\
       & = |\{w\mid (\varphi(u_{m,n}),w)\in R\}|\\
       &  = |\{w\mid (\bar a^m a^n,w)\in R\}|\,.
  \end{align*}

  Let $\cA$ be a finite deterministic automaton
  accepting~$R^\otimes$. For $q$ a state of $\cA$ and $m\in\bN$, let
  $l_q(m)$ denote the number of paths from an initial state to $q$
  labeled $\bar a^m\otimes w'$ for some $w'\in\{a,\bar
  a\}^m$. Similarly, let $r_q(n)$ denote the number of paths from $q$
  to some final state labeled $a^n\otimes w''$ for some
  $w''\in\{a,\bar a\}^n$. Then, for $m,n\in \bN$, we have
  \[
    \min(m,n)+1=\sum_{q\in Q}l_q(m)\cdot r_q(n)
  \]
  since the sum equals the number of words $\bar a^m a^n\otimes w\in
  R^\otimes$.

  Since $\bN^Q\times\bN^Q$, ordered componentwise, is a well-partial
  order, there are $m<n$ with $l_q(m)\le l_q(n)$ and $r_q(m)\le
  r_q(n)$ for all $q\in Q$.  Note that
  \[
    \sum_{q\in Q}l_q(m)\cdot r_q(m) = \min(m,m)+1 
     < \min(n,n)+1 = \sum_{q\in Q}l_q(n)\cdot r_q(n)\,.
  \]
  Hence there is $q\in Q$ with $l_q(m)<l_q(n)$ or
  $r_q(m)<r_q(n)$. Assuming the former, we get
  \begin{align*}
    m+1  = \min(m,m)+1 &=\sum l_q(m)\cdot r_q(m) \\
        &< \sum l_q(n)\cdot  r_q(m) = \min(n,m)+1=m+1\,,
  \end{align*}
  a contradiction. In the latter case, we similarly get
  \begin{align*}
    m+1 = \min(m,m)+1&=\sum l_q(m)\cdot r_q(m)\\
    & < \sum l_q(m)\cdot r_q(n) = \min(m,n)+1=m+1\,,
  \end{align*}
  again a contradiction.\qed
\end{proof}

\noindent
Recently, the notion of an automatic group has been extended to that
of Cayley graph automatic groups \cite{KhKhMi2011}. This notion can
easily be extended to monoids. It is not clear whether the monoid of
queue actions is Cayley graph automatic.

Note that $\cQ$ is not automatic in the sense of Khoussainov and
Nerode~\cite{KhoN95}: This is due to the fact that $\nat(A^*)$ is
isomorphic to $A^*$ and an element of $\cQ$ is in $\nat(A^*)$ if and
only if it cannot be written as $r\bar{a}s$ for $r,s\in\cQ$ and $a\in
A$.  Hence, using the $\bar{a}$ for $a\in A$ as parameters, $A^*$ is
interpretable in first order logic in $\cQ$. Therefore, since $A^*$ is
not automatic in this sense \cite{BluG04}, neither is $\cQ$~\cite{KhoN95}.

\bibliography{bibliography}\bibliographystyle{alpha}
\end{document}